\documentclass[12pt,letterpaper]{article}

\usepackage{amsmath,amssymb,calc}
\usepackage{graphicx}

\newcommand\be{\begin{equation}}
\newcommand\ee{\end{equation}}
\newcommand{\bea}{\begin{eqnarray}}
\newcommand{\eea}{\end{eqnarray}}

\newcommand{\nn}{\nonumber}
\newcommand{\pd}{\partial}

\def\id{\protect{{1 \kern-.28em {\rm l}}}}

\def\ve{\varepsilon}

\def\id{\protect{{1 \kern-.28em {\rm l}}}}

\let\non\nonumber

\begin{document}

\begin{titlepage}
\begin{center}
\hfill \\
\vspace{2cm}
{\Large {\bf Systematics of Constant Roll Inflation
\\[3mm] }}

\vskip 1.5cm
{\bf Lilia Anguelova${}^a$\footnote{anguelova@inrne.bas.bg}, Peter Suranyi${}^b$\footnote{peter.suranyi@gmail.com} and L.C.R. Wijewardhana${}^b$\footnote{rohana.wijewardhana@gmail.com}\\
\vskip 0.5cm  {\it ${}^a$ Institute for Nuclear Research and Nuclear Energy, BAS, Sofia, Bulgaria\\ ${}^b$ Department of Physics, University of Cincinnati,
Cincinnati, OH 45221, USA}\non\\}

\vskip 6mm

\end{center}

\vskip .1in
\vspace{1cm}

\begin{center} {\bf Abstract}\end{center}

\vspace{-1cm}

\begin{quotation}\noindent

We study constant roll inflation systematically. This is a regime, in which the slow roll approximation can be violated. It has long been thought that this approximation is necessary for agreement with observations. However, recently it was understood that there can be inflationary models with a constant, and not necessarily small, rate of roll that are both stable {\it and} compatible with the observational constraint $n_s \approx 1$. We investigate systematically the condition for such a constant-roll regime. In the process, we find a whole new class of inflationary models, in addition to the known solutions. We show that the new models are stable under scalar perturbations. Finally, we find a part of their parameter space, in which they produce a nearly scale-invariant scalar power spectrum, as needed for observational viability.

\end{quotation}
\vfill

\end{titlepage}

\eject

\tableofcontents

\section{Introduction}

It is well-known that on large scales the present day Universe is quite homogeneous and isotropic. The most promising candidate for explaining these properties is the inflationary scenario, according to which in the early Universe there was a period of accelerated expansion of space itself. An additional advantage of inflation is that it naturally generates the density perturbations that become the seeds of the subsequent large-scale structure of the Universe. Conventional inflationary models assume that this accelerated expansion is due to a fundamental scalar $\phi$ called inflaton, whose potential energy dominates the energy density of the Universe during the inflationary period. A standard approximation used in these models is to neglect the second time-derivative of the inflaton, $\ddot{\phi}$\,, or more precisely to assume that $\frac{\ddot{\phi}}{H \dot{\phi}} <\!\!< 1$\,, where $H$ is the Hubble parameter. This so called slow-roll approximation is widely thought to be necessary for agreement with observations, since it leads to a nearly scale-invariant spectrum of density perturbations. At present, it is indeed well-established observationally \cite{CMB} that the scalar spectral index is $n_s \approx 1$ and thus the spectrum is almost scale invariant. 

It has long been known, however, that there is an exception to the expectation that only slow roll can give $n_s=1$\,. $\!\!$Namely, a so called ultra-slow roll \cite{TW} stage can also produce a scale invariant spectrum \cite{WK}, although in that case $\frac{\ddot{\phi}}{H \dot{\phi}} = - 3$\,. The reason this kind of inflationary regime has attracted less interest in the past (although, see \cite{ultra-slow-roll}) is that it is unstable. In other words, it can last only for a few e-folds \cite{TW} and thus cannot constitute a full-fledged inflationary model by itself. Nevertheless, recently it was realized that a brief non-slow roll stage around the time of horizon exit of the largest observed CMB scales, followed by standard slow-roll inflation, may help explain the low multipole-moment anomaly in the CMB spectrum \cite{CK}. So even unstable inflationary regimes can be important for phenomenology.

Furthermore, it was shown recently that it is possible to have a {\it stable} non-slow roll regime, which still produces a nearly scale invariant spectrum \cite{MSY}. This work studied a class of models defined by a constant rate of roll $\frac{\ddot{\phi}}{H \dot{\phi}} = const$\,, which is obviously a generalization of the ultra-slow roll case. They found that for certain values of that constant one can have phenomenologically viable inflationary models. Further comparison of those models to observations was carried out in \cite{MS}. Since then, the constant roll regime has attracted a lot of interest; see, for example, the investigation of transitions between stages with different rates of roll in \cite{OO,VO} or the studies of constant roll in modified theories of gravity in \cite{NOO,MS2,VO2,OOS,AHNOO}. Given the increasing (and quite impressive) precision of cosmological observation nowadays, it is indeed imperative to understand better the full set of theoretical models that can lead to a nearly-scale invariant spectrum of perturbations. 

In the present paper we will study the condition for a constant rate of roll systematically. In the process, we will obtain the solutions of \cite{MSY} in a more straightforward manner. More importantly, we will find a new class of constant-roll inflationary models that are both stable under perturbations and compatible with the observational requirement that $n_s \approx 1$. The accelerated expansion in these models can last only for a finite period of time, although the duration of that period can be made arbitrarily large by a suitable choice of an integration constant. So, to embed such models within a realistic description of the evolution of the Universe, one needs to assume that they are valid only until a certain moment of time. After that, a different effective description, which is beyond the scope of the current paper, has to take over. Such an assumption, however, is rather common in inflation model building, including in \cite{MSY,MS}, since the transition from the inflationary stage to the subsequent hot Big Bang is, in itself, a separate quite nontrivial topic of research.

The organization of this paper is the following. In Section \ref{SecCRR}, we write the constant roll condition as an ODE for the Hubble parameter, as a function of time, and find all of its solutions. Then we compute the scale factors and inflaton solutions for each of these cases. In Section \ref{SecNM}, we show that among them there is a new class of inflationary models, namely solutions that admit a period of positive acceleration. We then find the inflaton potential for this class of models. In Section \ref{SecPertSt}, we show that these new models are stable under scalar perturbations and that there is a part of their parameter space, in which they give a scalar spectral index $n_s \approx 1$\,. In Section \ref{SecDisc}, we summarize this work and discuss open issues for the future. Finally, in the two Appendices we give technical details necessary for the considerations in Section \ref{SecCRR}.

\section{Constant roll regime} \label{SecCRR}

Our goal will be to investigate systematically a certain kind of non-standard inflationary solutions to the usual equations of motion. Recall that the action for a single scalar field $\phi$ minimally coupled to gravity is:
\be
S = \int d^4 x \sqrt{-g} \left[ \frac{R}{2} + \frac{1}{2} g^{\mu \nu} \pd_{\mu} \phi \pd_{\nu} \phi - V(\phi) \right] \,\,\, ,
\ee
where $V(\phi)$ is the scalar potential. With the usual metric ansatz
\be
ds_4^2 = -dt^2 + a^2 (t)\,d\vec{x}^2 \,\,\, ,
\ee 
where $a(t)$ is the scale factor, the equations of motion are the Friedman equations:
\bea \label{FriedEq}
3 H^2 &=& \frac{\dot{\phi}^2}{2} + V \,\,\, , \nn \\
- 2 \dot{H} &=& \dot{\phi}^2 \,\,\, ,
\eea
as well as the field equation for the scalar:
\be \label{EoMphi}
\ddot{\phi} + 3 H \dot{\phi} + \frac{\pd V}{\pd \phi} = 0 \,\,\, .
\ee
As usual, $H$ in (\ref{FriedEq})-(\ref{EoMphi}) is the Hubble parameter defined as: 
\be \label{Hdaa}
H(t) = \frac{\dot{a}(t)}{a(t)} \,\,\, .
\ee

The standard way of finding an inflationary solution to the above system of field equations is to assume the slow roll approximation. That means assuming that the slow roll parameters \cite{LPB,DB}:
\be \label{eta_phi}
\ve = - \frac{\dot{H}}{H^2} \qquad {\rm and} \qquad \eta = -\frac{\ddot{\phi}}{H \dot{\phi}}
\ee
satisfy the conditions
\be
\ve <\!\!< 1 \qquad {\rm and} \qquad \eta <\!\!< 1 \,\,\, .
\ee
In that case, the scale factor increases with time almost exponentially and the inflationary background is well-approximated by de Sitter space. 

Here we will be interested in a different class of inflationary solutions, for which the slow roll parameters are not necessarily small. Namely, we will study the so called regime of constant roll, defined by \cite{MSY}:
\be \label{eta_const_roll}
\eta = -\frac{\ddot{\phi}}{H \dot{\phi}} \equiv const \,\,\, .
\ee
For that purpose, let us note that the $\eta$ parameter can also be written as \cite{OO}:
\be \label{eta_H}
\eta = - \frac{\ddot{H}}{2 H \dot{H}} \,\,\, .
\ee
One can easily see that the definitions of $\eta$ in (\ref{eta_phi}) and (\ref{eta_H}) are equivalent, whenever the second equation in (\ref{FriedEq}) is satisfied. Hence we can rewrite the constant roll condition (\ref{eta_const_roll}) as: 
\be \label{Heq}
- \frac{\ddot{H}}{2 H \dot{H}} = const \equiv c \, .
\ee

Clearly, (\ref{Heq}) is an ODE for the function $H(t)$, that can be solved easily. Once we have $H(t)$, it is straightforward to obtain all the other functions, important for any inflationary model. Namely, $a(t)$ can obviously be found from solving (\ref{Hdaa}), while the inflaton can be determined, due to the second equation in (\ref{FriedEq}), via:
\be \label{InflSol}
\phi (t) = \pm \int \sqrt{-2 \dot{H}} \, dt \,\,\, .
\ee
And, finally, the potential $V(\phi)$ can be found from the first equation in (\ref{FriedEq}), after inverting the inflaton to obtain the function $t = t(\phi)$ .\footnote{Of course, for that purpose we assume that $\dot{\phi} \neq 0$.} Notice that, once the Friedman equations (\ref{FriedEq}) have been solved, the equation of motion of the inflaton (\ref{EoMphi}) is automatically satisfied as well. Indeed, viewing $V$, $H$ and $\dot{\phi}$ as functions of $\phi$ via $V(t(\phi))$, $H(t(\phi))$ and $\dot{\phi} (t(\phi))$, we obtain upon differentiating the first Friedman equation:
\be \label{Vdphi}
\pd_{\phi} V = 6 H \pd_{\phi} H - \dot{\phi} \pd_{\phi} \dot{\phi} = 6 H \frac{\dot{H}}{\dot{\phi}} - \ddot{\phi} = - 3 H \dot{\phi} - \ddot{\phi} \,\,\, ,
\ee
where in the last step we have used the second Friedman equation $\dot{\phi}^2 = - 2 \dot{H}$. Clearly, the relation (\ref{Vdphi}) is exactly the same as the field equation (\ref{EoMphi}). So, in view of all of the above, we can view (\ref{Heq}) as the master equation that determines an inflationary model in the constant roll regime. Let us now turn to solving it.

\subsection{Hubble parameter}

In order to investigate equation (\ref{Heq}) systematically, let us first note that it does not depend explicitly on $t$. So we can simplify it by introducing the notation:
\be \label{HdU}
\dot{H} \equiv U(H) \,\, .
\ee
In terms of the new function $U(H)$, one has $\ddot{H} = U \partial_H U$. Therefore, (\ref{Heq}) acquires the form:
\be
- \frac{\partial_H U}{2 H}  = c \,\, .
\ee
This has as a general solution
\be \label{UH}
U(H) = - c H^2 + C_1 \,\, ,
\ee
where $C_1$ is an integration constant. Now, recalling that $U = \dot{H}$ and integrating (\ref{UH}), we find:
\be
t+C_2 = \int \frac{dH}{C_1 - c H^2}
\ee
with $C_2$ being another integration constant. To perform explicitly the last integral, we need to make assumptions about the signs of the integration constants. First, note the special choice $C_1 = 0$, which gives:
\be \label{H1t}
H_{(0)} (t) = \frac{1}{c \,t + C_2} \,\, .
\ee 
Now, keeping $C_1 \neq 0$ and considering different choices for the signs of $c$ and $C_1$, we find the solutions:
\bea \label{Hsol}
H_{(1)} (t) &=& \frac{\sqrt{C_1 c}}{c} \,\coth \!\left( \!\sqrt{C_1 c} \,(t+C_2) \!\right) \qquad {\rm for} \qquad sgn(C_1)=sgn(c) \,\,\, , \\
H_{(2)} (t) &=& \frac{\sqrt{C_1 c}}{c} \,\tanh \!\left( \!\sqrt{C_1 c} \,(t+C_2) \!\right) \qquad {\rm for} \qquad sgn(C_1)=sgn(c) \,\,\, , \nn \\
H_{(3)} (t) &=& \frac{\sqrt{|C_1 c|}}{c} \,\cot \!\left( \!\sqrt{|C_1 c|} \,(t+C_2) \!\right) \qquad {\rm for} \qquad sgn(C_1)=-sgn(c) \,\,\, , \nn \\
H_{(4)} (t) &=& - \,\frac{\sqrt{|C_1 c|}}{c} \,\tan \!\left( \!\sqrt{|C_1 c|} \,(t+C_2) \!\right) \qquad {\rm for} \qquad sgn(C_1)=-sgn(c) \,\,\, . \nn
\eea 

As pointed out in \cite{MSY}, the solution $H_{(0)}$ in (\ref{H1t}) gives power-law inflation \cite{AW,LM}\footnote{Note that there is a typo in equation (17) of \cite{MSY}. Namely, the expression there should be $H = \frac{1}{(3+\alpha) t}$\,.}, which is known to produce too large an amount of primordial gravity waves, to be viable according to present day observations. So just as there, we will not discuss this solution any further. As for (\ref{Hsol}), the solutions $H_{(1)}$, $H_{(2)}$ and $H_{(4)}$ with $C_2 = 0$ were written down in \cite{MSY}, while $H_{(3)}$ was omitted entirely. Note, however, that the presence of a suitable $C_2\neq 0$ may be physically useful, as it can regulate singular behavior in $H_{(1)}$ (at early times) and in $H_{(2)}$ (at late times), as will become more clear later on. Also and more importantly, \cite{MSY} only considered $H_{(1)}$ and $H_{(2)}$ as inflationary solutions. We will show in the following that $H_{(3)}$ and $H_{(4)}$ can lead to a constant-roll inflationary regime too. Although, as we will see, one obtains the same class of models from both $H_{(3)}$ and $H_{(4)}$, it will become clear that, in fact, the form $H_{(3)} (t)$ is more convenient for describing it.

Note that the solutions (\ref{Hsol}) can be written in a unified manner in the following way:
\be \label{HNewSol}
H (t) = h \, \frac{k \,e^{hct} + e^{-hct}}{k \,e^{hct}-e^{-hct}} \,\,\, ,
\ee
where the new integration constants $h$ and $k$ are complex numbers. Indeed, for example $H_{(1)}$ can be obtained by taking both $h$ and $k$ to be real. In that case and assuming that $k>0$, one can rewrite (\ref{HNewSol}) as:
\be \label{Hcothhk}
H (t) = h \, \frac{e^{hct + \frac{1}{2} \ln k} + e^{-hct - \frac{1}{2} \ln k}}{e^{hct + \frac{1}{2} \ln k}-e^{-hct - \frac{1}{2} \ln k}} = \,h \,\coth \!\left( hct + \frac{1}{2} \ln \!k \right) \,\,\,\, {\rm for} \,\,\,\,\, h \in \mathbb{R} \,\, , \,\, k>0 \,\,\, .
\ee
Similarly, (\ref{HNewSol}) gives:
\bea \label{Hhkcases}
H (t) &=& \,h \,\tanh \!\left( hct + \frac{1}{2} \ln \!|k| \right) \quad {\rm for} \quad h \in \mathbb{R} \,\, , \,\, k<0 \,\,\, , \nn \\
H (t) &=& \,\hat{h} \,\cot \!\left( \hat{h}ct + \frac{\theta}{2} \right) \quad {\rm for} \quad h= i \hat{h} \,\, {\rm and} \,\, \hat{h} \in \mathbb{R} \,\,\, , \,\,\, k=e^{i \theta} \,\,\, , \nn \\
H (t) &=& - \,\hat{h} \,\tan \!\left( \hat{h}ct + \frac{\theta}{2} \right) \quad {\rm for} \quad h= i \hat{h} \,\, {\rm and} \,\, \hat{h} \in \mathbb{R} \,\,\, , \,\,\, k = - e^{i \theta} \,\,\, .
\eea
Clearly, (\ref{Hcothhk})-(\ref{Hhkcases}) are true for either sign of $c$. Note also that, obviously, having $C_2 \neq 0$ in (\ref{Hsol}) corresponds to having $k \neq 1$ in (\ref{HNewSol}). 

Mathematically, (\ref{HNewSol}) is a solution of (\ref{Heq}) for any complex values of $h$ and $k$. However, physically we have to take only values, such that the Hubble parameter is real. One can show that $H (t)$ in (\ref{HNewSol}) is a real function of time only for the above choices of $h$ and $k$. More precisely, both $h$ and $k$ have to be real or both have to be complex and of the type: $h = i \mathbb{R}$ and $k = \pm e^{i \theta}$. For the proof see Appendix \ref{HComplexConsts}. To summarize, the only real functions one obtains from (\ref{HNewSol}) are the solutions in (\ref{Hsol}). Hence the unifying form (\ref{HNewSol}), restricted to real values, is precisely equivalent to the set of solutions (\ref{Hsol}). Finally, note that the following choices for the pairs of $h$ and $k$ in (\ref{HNewSol}):
\bea \label{hk4cases}
(h,k) &=& (h \in \mathbb{R}\,,\,k>0)_{(1)} \quad , \quad (h \in \mathbb{R}\,,\,k<0)_{(2)} \quad , \nn \\
&&(h=i \mathbb{R}\,,\,k=e^{i\theta})_{(3)} \quad , \quad (h=i \mathbb{R}\,,\,k=-e^{i\theta})_{(4)} 
\eea
give the corresponding Hubble parameters $H_{(i)} (t)$ of (\ref{Hsol}), upon an obvious redefinition of the integration constants.

\subsection{Scale factor}

Let us now compute the scale factor $a(t)$. Substituting (\ref{HNewSol}) in (\ref{Hdaa}), we find:
\be \label{sf}
a (t) = C_a \left( k e^{h c t} - e^{- h c t} \right)^{1/c} \,\,\, ,
\ee
where $C_a$ is an integration constant. For the choices of pairs $(h,k)_{(i)}$ in (\ref{hk4cases}), the expression (\ref{sf}) acquires the respective forms:
\bea \label{scalefactors}
a_{(1)} (t) &=& C^a_1 \,\sinh^{1/c} \!\left( \!\sqrt{C_1 c} \, (t + C_2) \!\right) \,\,\, , \nn \\
a_{(2)} (t) &=& C^a_2 \cosh^{1/c} \!\left( \!\sqrt{C_1 c} \, (t+C_2) \right) \,\,\, , \nn \\
a_{(3)} (t) &=& C^a_3 \,\sin^{1/c} \!\left( \!\sqrt{|C_1 c|} \, (t + C_2) \!\right) \,\,\, , \nn \\
a_{(4)} (t) &=& C^a_4 \,\cos^{1/c} \!\left( \!\sqrt{|C_1 c|} \, (t + C_2) \!\right) \,\,\, , \label{accel}
\eea
upon suitably redefining the overall constant $C_a$ to absorb an (in general, complex) numerical factor. One can easily verify that, solving (\ref{Hdaa}) directly for each of the functions in (\ref{Hsol}), gives exactly the same expressions as in (\ref{scalefactors}), as should be the case. 

For future use, let us also write down the acceleration that follows from (\ref{sf}):
\be \label{dda}
\ddot{a} (t) = C_a \,h^2 \left( k e^{hct} - e^{-hct} \right)^{\frac{1-2c}{c}} \,\left[ \left( k e^{hct} + e^{-hct} \right)^2 - 4 c k \right] \, .
\ee
Recall that the definition of inflation is a period of time during which $\ddot{a} (t) > 0$. Clearly, this condition will be satisfied for some values of the various constants in (\ref{dda}) and not for others. In particular, it might seem at first sight that the cases with $a_{(3)} (t)$ and $a_{(4)} (t)$ in (\ref{scalefactors}) can give only $\ddot{a} (t) < 0$, as stated in \cite{MSY}. We will show in the next section that this is not the case. Namely, it will turn out that, for $0<c<1$, one can have periods of time during which $\ddot{a}_{(3),(4)} > 0$. 

Due to the oscillatory nature of the scale factor in these cases, the intervals of time with $\ddot{a}_{(3),(4)} > 0$ will be finite, unlike in the cases with $a_{(1)}$ and $a_{(2)}$. This is not a real deficiency, however, since the length of any time-interval with positive acceleration can be extended arbitrarily by choosing suitably the value of the integration constant $C_1$. It is also important to remember that inflation in the Early Universe {\it has} to last only for a (rather) short period of time, after which one has to assume that some other mechanism is starting to play a dominant role and the inflationary stage is exited. Hence, a priori, there is no reason to consider the class of models with scale factor $a_{(3),(4)}$ any less seriously than the models with $a_{(1)}$ and $a_{(2)}$, that were investigated in detail in \cite{MSY}. In addition, any finite (even if very short) period of constant-roll inflation might be useful in understanding the observed low-$l$ anomaly of the CMB \cite{CMB}. Indeed, it has been argued recently \cite{CK} that this anomaly could be explained, if the standard slow-roll inflationary expansion is preceded by a brief non-slow roll stage, as mentioned in the introduction.

\subsection{Inflaton solution} \label{SecInflaton}

Recall that the inflaton is determined from the Hubble parameter via (\ref{InflSol}):
\be \label{infl}
\phi (t) = \mp \int \sqrt{-2 \dot{H}} \, dt \,\,\, .
\ee
Obviously, to have a real inflaton solution (and, in fact, inflation), we need $\dot{H} < 0$. This condition leads to important restrictions on the parameter space of each of the models with Hubble parameters $H_{(1)}$,...,$H_{(4)}$.

Indeed, from (\ref{HNewSol}) we have:
\be \label{dHcksigns}
\dot{H} (t) = - \frac{4 \,c \,k \,h^2}{( ke^{hct}-e^{-hct})^2} \,\,\, .
\ee
This immediately implies that, for real $h$, the parameter $c$ has to have the same sign as the integration constant $k$. In view of (\ref{hk4cases}), that means that the case with Hubble parameter $H_{(1)} (t)$ can only have $c>0$, while the case with $H_{(2)} (t)$ has to have $c<0$. As a result, to have positive Hubble parameter, one needs to take the following time ranges: $(t+C_2) \in [0,\infty)$ for $H_{(1)}$ and $(t+C_2) \in (-\infty , 0\,]$ for $H_{(2)}$. To summarize, in order to obtain inflationary models in the first two cases, we have to take:
\bea \label{H1H2constr}
&&H_{(1)} (t) \,\,\, : \quad c>0 \quad {\rm and} \quad (t+C_2) \in [\,0,\infty) \quad , \nn \\
&&H_{(2)} (t) \,\,\, : \quad c<0 \quad {\rm and} \quad (t+C_2) \in (-\infty , 0\,] \quad .
\eea
These restrictions are in perfect agreement with the considerations of \cite{MSY}, although they were not discussed so explicitly there. Note also that, for $C_2 = 0$\,, the Hubble parameters $H_{(1),(2)}$ exhibit singular behavior when $t \rightarrow 0$\,, as can be seen from (\ref{Hsol}). More precisely, at early times $H_{(1)} \rightarrow \infty$\,, whereas at late times $H_{(2)} \rightarrow 0$\,. Clearly, taking a suitable $C_2\neq 0$ can regulate this behavior.

In the remaining two cases in (\ref{hk4cases}), with $h$ purely imaginary, we obtain from (\ref{dHcksigns}):
\be
\dot{H}_{(3)} = - \frac{c \hat{h}^2}{\sin^2 (\hat{h} ct + \frac{\theta}{2})} \qquad {\rm and} \qquad \dot{H}_{(4)} = - \frac{c \hat{h}^2}{\cos^2 (\hat{h} ct + \frac{\theta}{2})} \quad ,
\ee
where we have substituted the corresponding pairs $(h,k)_{(i)}$ from (\ref{hk4cases}), together with the notation $h = i \hat{h}$ and $\hat{h} \in \mathbb{R}$. Obviously, now $c$ has to be positive in both cases to ensure $\dot{H}_{(3),(4)} < 0$. This condition, in fact, implies that both cases $(3)$ and $(4)$ actually give the same class of models. To make that more explicit, let us first introduce for convenience the notation:
\be \label{varphi_def}
\varphi \equiv \sqrt{|C_1| c} \,(t+C_2) \,\,\, . 
\ee
Now, to have a positive Hubble parameter in case $(3)$, given that $c>0$, one needs to consider either the range $\varphi \in [0, \frac{\pi}{2}]$ or the range $\varphi \in [\pi, \frac{3 \pi}{2}]$; see (\ref{Hsol})\,.{}\footnote{Note that the scale factor $a_{(3)} (\varphi)$ can be positive in the interval $[\pi, \frac{3 \pi}{2}]$ as well, even though $\sin \varphi < 0$ there, because we can always redefine the arbitrary integration constant $C_3^a$ and/or choose the parameter $c$ suitably, like for example by taking $c=\frac{1}{2m}$ with $m$ being a positive integer. \label{ftn}} Note, though, that $H_{(3)} (t)$ is the same function in both ranges, since $\cot (\varphi + \pi) = \cot (\varphi)$. So, without any loss of generality, we can write:
\be \label{H3range}
H_{(3)} (t) \,\,\, : \quad c>0 \quad {\rm and} \quad \varphi \in \left[ 0, \frac{\pi}{2} \right] \,\,\, .
\ee
Similarly, to have $H_{(4)} (t) > 0$, given that $c>0$, we need to take either $\varphi \in [\frac{\pi}{2}, \pi]$ or $\varphi \in [\frac{3 \pi}{2}, 2 \pi]$\,.{}\footnote{Clearly, the same remark as in footnote \ref{ftn} applies to $a_{(4)} (\varphi)$ in the interval $[\frac{\pi}{2}, \pi]$.} Again, $H_{(4)} (t)$ is the same function in both intervals, for the same reason as for $H_{(3)}$. Hence, in this case:
\be \label{H4range}
H_{(4)} (t) \,\,\, : \quad c>0 \quad {\rm and} \quad \varphi \in \left[ \frac{\pi}{2}, \pi \right] \,\,\, .
\ee
Notice that the intervals in (\ref{H3range}) and (\ref{H4range}) are shifted by $\frac{\pi}{2}$ compared to each other. So, due to $\tan (\varphi + \frac{\pi}{2}) = - \cot(\varphi)$, we actually have the same Hubble parameter in both cases $(3)$ and $(4)$. Therefore, since all functions of interest are derived from the Hubble parameter (as already explained), we can conclude that one obtains the {\it same} class of models from both cases $(3)$ and $(4)$. This is obvious for the scale factors $a_{(3)}$ and $a_{(4)}$ in (\ref{scalefactors}). It will turn out, though, to be a bit more non-trivial for the inflaton solutions, as we will see shortly. 

Now, for future use and for comparison with \cite{MSY}, let us write down the inflaton solutions in all four cases:
\bea \label{phicases}
\phi_{(1)} (t) &=& \pm \sqrt{\frac{2}{c}} \,\ln \!\left[ \,\coth \!\left( \frac{\sqrt{C_1 c}}{2} \,(t+C_2) \right) \right] + C_1^{\phi} \quad {\rm with} \quad c > 0 \,\,\, , \\
\phi_{(2)} (t) &=& \pm \sqrt{\frac{8}{|c|}} \,\arctan \!\left( e^{\sqrt{C_1 c} \,(t+C_2)} \right) + C_2^{\phi} \quad {\rm with} \quad c < 0 \,\,\, , \nn \\
\phi_{(3)} (t) &=& \pm \sqrt{\frac{2}{c}} \,\ln \!\left[ \cot \!\left( \frac{\sqrt{|C_1\!| c}}{2} \,(t+C_2) \right) \right] + C_3^{\phi} \quad {\rm with} \quad c > 0 \,\,\, , \nn \\
\phi_{(4)} (t) &=& \mp \sqrt{\frac{8}{c}} \,\,{\rm arctanh}\!\left[ \tan \!\left( \frac{\sqrt{|C_1\!| c} \,(t+C_2)}{2} \right) \right] + C_4^{\phi} \quad {\rm with} \quad c > 0 \,\,\, . \nn
\eea
where in $\phi_{(i)}$, $i=1,3,4$ we have assumed that $(t+C_2) > 0$ and in $\phi_{(2)}$ that $(t+C_2) < 0$. Note that the integration constants $C^{\phi}_i$ do not have to be real numbers, although of course we want the inflaton to be real. This observation will be important later on.

The expressions in (\ref{phicases}) were obtained directly by using each of (\ref{Hsol}) in (\ref{infl}). However, one should also be able to understand them as special cases, following from the unifying form of the Hubble parameter (\ref{HNewSol}). As this comparison is instructive, we will consider it in more detail here. Substituting (\ref{HNewSol}) in (\ref{infl}), we find:
\be \label{Infl_Sol}
\phi (t) = \pm \sqrt{\frac{8}{c}} \,\, {\rm arctanh} \!\left( \sqrt{k} e^{hct} \right) + C_{\phi} \,\,\, ,
\ee
where $C_{\phi}$ is an integration constant. To compare this with $\phi_{(1)}$, let us first note the obvious identifications between the pairs of integration constants $(h,k)$ and $(C_1,C_2)$, that can be read off from comparing $H_{(1)}$ in (\ref{Hsol}) with the expression (\ref{Hcothhk}), namely \,$h = \frac{\sqrt{C_1 c}}{c}$ \,and \,$\frac{1}{2} \ln k = \sqrt{C_1 c} \,C_2$\,. This, together with $c>0$, immediately implies that $h>0$ in case $(1)$. Note also, that if we want $H_{(1)} (t) > 0$ for $\forall t \in [0,\infty)$, then we must have $C_2\ge 0$. Hence in this case $k\ge 1$. Therefore, the argument of the {\it arctanh} in (\ref{Infl_Sol}) is always $\ge 1$ for any positive $t$, implying that the function ${\rm arctanh} (\sqrt{k} e^{hct})$ is complex. However, its imaginary part is constant (see (\ref{Imarctanh_zge1})) and thus it can be canceled by an appropriate imaginary part of the integration constant $C_{\phi}$. The real part, then, is ${\rm arccoth} (\sqrt{k} e^{hct})$; see (\ref{Rearctanh}). Now, using that \,$2 \,{\rm arccoth} (e^{2x}) = \ln \left( \frac{e^{2x} + 1}{e^{2x} - 1} \right) = \ln (\coth (x))$ \,for any $x$, we see that (\ref{Infl_Sol}) becomes exactly $\phi_{(1)}$ in (\ref{phicases}). 

Identifying $\phi_{(2)}$ as the appropriate special case of (\ref{Infl_Sol}), is even more direct. Namely, recall that solution $(2)$ is obtained for $k < 0$ and $c<0$, as explained above. Substituting $k = - |k|$ and $c = - |c|$ in (\ref{Infl_Sol}) and recalling that \,${\rm arctanh} (ix) = i \arctan (x)$ \,for any $x$, we see that (\ref{Infl_Sol}) becomes exactly $\phi_{(2)}$, upon setting \,$h = \frac{\sqrt{C_1 c}}{c}$ \,and \,$\frac{1}{2} \ln |k| = \sqrt{C_1 c} \,C_2$\,. The latter identifications can be immediately read off from comparing $H_{(2)}$ in (\ref{Hsol}) with the first line of (\ref{Hhkcases}). 

Finally, in both cases $(3)$ and $(4)$, the comparison between (\ref{Hsol}) and (\ref{Hhkcases}) leads to the same identification between the pairs of integration constants: $\hat{h} = \frac{\sqrt{|C_1\!| c}}{c}$ and $\frac{\theta}{2} = \sqrt{|C_1\!| c} \,C_2$\,. Now, substituting $h = i \hat{h}$, with $\hat{h} \in \mathbb{R}$, and $ k = e^{i \theta}$ in (\ref{Infl_Sol}), the latter acquires the form:
\be
{\rm case\,\,(3)} \,\,\, : \quad \phi (t) = \pm \sqrt{\frac{8}{c}} \,\,{\rm arctanh} (e^{i \varphi}) + C_{\phi} \,\,\, ,
\ee
where $\varphi$ is the same as in (\ref{varphi_def}). Then, using (\ref{arctanhphase_final}), we find exactly the expression for $\phi_{(3)}$ in (\ref{phicases}). Similarly, substituting $h = i \hat{h}$ and $ k = - e^{i \theta}$ in (\ref{Infl_Sol}), we obtain:
\be \label{case4IexpIVarphi}
{\rm case\,\,(4)} \,\,\, : \quad \phi (t) = \pm \sqrt{\frac{8}{c}} \,\,{\rm arctanh} (i e^{i \varphi}) + C_{\phi} \,\,\, , 
\ee
which upon using (\ref{arctanhiphase_final}) gives precisely $\phi_{(4)}$ in (\ref{phicases}). The latter expression agrees with the inflaton solution in \cite{MSY} for the corresponding case.

Note that, from the forms of $\phi_{(3)}$ and $\phi_{(4)}$ in (\ref{phicases}), it is not immediately clear that they give the same function, up to an overall $\pm$\,, in the intervals in (\ref{H3range}) and (\ref{H4range}). For completeness, we show in Appendix \ref{ApExpIphi} that this is indeed the case, as expected. On the other hand, the general expression (\ref{Infl_Sol}) makes this very easy to understand. Indeed, recall that the ranges of the variable are: $\varphi_{(3)} \in [0,\frac{\pi}{2}]$ for case $(3)$ and $\varphi_{(4)} \in [\frac{\pi}{2},\pi]$ for case $(4)$. Hence, from $\varphi_{(4)} = \varphi_{(3)} + \frac{\pi}{2}$\,, it immediately follows that ${\it arctanh} (i e^{i \varphi_{(4)}}) = - {\it arctanh} (e^{i \varphi_{(3)}})$\,, or in other words that $\phi_{(4)} = -\phi_{(3)}$\,. Given the overall $\pm$ in each of them, this proves that $\phi_{(3)} (t)$ and $\phi_{(4)} (t)$ are the same function.

\section{New class of models} \label{SecNM}
\setcounter{equation}{0}

In the previous section we showed that cases $(3)$ and $(4)$ give the same class of models. Now we will show that in these models one can have intervals of time with positive acceleration. In other words, we will see that this is a new class of constant-roll inflationary models. 

Let us begin by summarizing what we have found about cases $(3)$ and $(4)$ so far. The parameter $c$ has to satisfy the constraint
\be
c > 0 \,\,\, ,
\ee
in order to have $\dot{H} < 0$. Hence, setting $C_2=0$ and introducing for convenience the notation 
\be \label{Nthetahat}
N \equiv \sqrt{|C_1\!| c} \quad ,
\ee 
while keeping in mind that $N>0$ in the following, we take the interval:
\be \label{interval}
N t \in \left[ 0, \frac{\pi}{2} \right] \,\,\, ,
\ee
in order to ensure that the Hubble parameter is positive. We have seen that every other interval, that gives positive $H_{(3)}$ or $H_{(4)}$, leads to the same class of models. So we can restrict to (\ref{interval}) without any loss of generality. Then, the Hubble parameter, scale factor and inflaton of the new class of models, that we will be studying, have the form:
\be \label{NewModel_setup}
H_{(3)} = \frac{N}{c} \cot (Nt) \,\,\,\, , \,\,\,\, a_{(3)} = C_3^a \sin^{1/c} (N t) \,\,\,\, , \,\,\,\, \phi_{(3)} = \pm \sqrt{\frac{2}{c}} \ln \!\left[ \cot \!\left( \frac{Nt}{2} \right) \!\right] + C_3^{\phi} \,\,\, .
\ee
Note that, although (\ref{interval}) implies a finite interval for $t$, namely $t \in \left[ 0, \frac{\pi}{2 N} \right]$, obviously the length of this interval can be made arbitrarily large by choosing suitably the integration constant $N$:
\be \label{Nless1}
N <\!\!< 1 \,\,\, .
\ee

We will see shortly that requiring positive acceleration imposes further constraints on the model.

\subsection{Positive acceleration} \label{SecPosAccel}

Let us now investigate under what conditions one can have $\ddot{a}_{(3)} (t) > 0$. Differentiating the scale factor in (\ref{NewModel_setup}), we find:
\be
\ddot{a}_{(3)} (t) = C^a_3 \,\frac{N^2}{c^2} \left[ \sin \!\left( N t \right) \right]^{\frac{1}{c} - 2} \left[ \,\cos^2 \!\left( N t \right) \!- c \,\right] \,\,\, ,
\ee
which can be rewritten nicely as:
\be \label{accel5}
\ddot{a}_{(3)} (t) = \frac{N^2}{c^2} \frac{a_{(3)} (t)}{ \sin^2 ( N t ) } \left[ \,\cos^2 \!\left( N t \right) \!- c \,\right] \,\,\, .
\ee
Hence, the condition for positive acceleration is:
\be \label{Condaddg0}
\cos^2 \!\left( N t \right) \! \,> \,c \,\,\, .
\ee
Clearly, to be able to solve it, we have to take:
\be
c < 1 \,\,\, .
\ee
In addition, though, we need to restrict the interval for $Nt$ further, compared to (\ref{interval}). This still does not prevent us from having a large $t$-interval for the same reason as in the discussion above eq. (\ref{Nless1}). In fact, note that changing the upper bound of the interval in (\ref{interval}) just amounts to redefining the arbitrary constant $N$. Indeed, one can always redefine \,$N \,\rightarrow \,\hat{N} \equiv \frac{2}{\pi} \theta_* N$ \,with some fixed $\theta_* \!< \frac{\pi}{2}$\,. Then the interval $Nt \in [0, \frac{\pi}{2}]$ transforms to $\hat{N} t \in [0, \theta_*]$\,. So shortening the interval in (\ref{interval}) is part of our freedom to choose the integration constant $N$ .\footnote{At first sight, it might seem that this argument is specific to the interval $[0,\frac{\pi}{2}]$\,, as the other intervals, in which $H_{(3),(4)} > 0$\,, have a non-vanishing lower bound. If that were true, it would contradict the statement that all of those intervals lead to the same class of models. However, note that in writing (\ref{interval}) we have set $C_2 = 0$\,. Instead, in the interval $[\frac{\pi}{2},\pi]$\,, for instance, we can fix the second integration constant in the pair $(C_1,C_2)$ by setting $\sqrt{|C_1\!|\,c} \,C_2 = \frac{\pi}{2}$\,. Then, we have again $Nt \in [0,\frac{\pi}{2}]$ and so can use the same redefinition of $N$, as in the main text. Obviously, the same goes for the intervals $[\pi,\frac{3\pi}{2}]$ and $[\frac{3\pi}{2},2\pi]$\,, with the choices $\sqrt{|C_1\!|\,c} \,C_2 = \pi$ and $\sqrt{|C_1\!|\,c} \,C_2 = \frac{3\pi}{2}$\,, respectively.} 

Now let us consider condition (\ref{Condaddg0}). In the interval (\ref{interval}), its solution is given by $Nt < {\rm arccos} (\sqrt{c})$ .\footnote{We assume that the function {\it arccos} is restricted to its principal branch.} In other words, to ensure $\ddot{a}_{(3)} (t) > 0$ and thus an inflationary period, we need to take:
\be \label{Ntmax_c}
Nt \in \left[\,0, \,{\rm arccos (\sqrt{c})}\,\right) \,\,\, .
\ee
Of course, this is the maximal interval, that solves our constraint. However, we are free to choose any subinterval of it, as discussed above. This will be useful shortly.

So far, we have considered the condition to have positive acceleration. It is also worth discussing under what conditions the acceleration is increasing or decreasing. Note that in the pure de Sitter case, as well as the standard slow-roll models that by definition are small deviations from it, the acceleration is increasing with time. Here, however, that is not always the case. Indeed, let us compute the derivative of $\ddot{a}_{(3)}(t)$:
\be
\frac{d\ddot{a}_{(3)}}{dt} = \frac{N^2}{c^2} \frac{a_{(3)} H_{(3)}}{\sin^2 (N t)} \left[ \cos^2 (N t) - 3c + 2c^2 \right] \,\,\, .
\ee
Obviously, to have $\frac{d\ddot{a}_{(3)}}{dt} >0$, we need:
\be \label{Condaccelpos}
\cos^2 (N t) > 3c - 2c^2 \,\,\, .
\ee
Note, however, that for $\frac{1}{2} < c < 1$ we have $3c - 2c^2 > 1$ and thus an $\ddot{a}_{(3)}$ that is decreasing with time. Therefore, to have any interval with increasing $\ddot{a}_{(3)}$, we have to take:
\be
c < \frac{1}{2} \,\,\, .
\ee
Then the condition (\ref{Condaccelpos}) can be solved, giving:
\be
Nt < {\rm arccos} \left( \!\sqrt{3 c - 2 c^2} \right) \,\,\, .
\ee
Now, since for $\forall c \in (0,\frac{1}{2})$ we have $3c-2c^2 > c$, it follows that:
\be
{\rm arccos} \left( \sqrt{3c-2c^2} \right) < \,{\rm arccos} \left( \sqrt{c} \right) \,\,\, .
\ee
Therefore, if we consider the entire interval (\ref{Ntmax_c}), at first $\ddot{a}_{(3)}$ will be increasing with time. Then, after the moment $t_* = \frac{1}{N} \arccos \left( \sqrt{3c-2c^2} \right)$, it will start decreasing. Finally, if we want to consider models with a more usual (in the sense of de Sitter-like) behavior, we can choose the end-point of the $Nt$-interval to be $Nt_*$, or even any point inside $[0, Nt_*)$ independent of $c$. For future convenience, we can take for instance:
\be \label{NtintPio4}
Nt \in \left[ 0, \frac{\pi}{4} \right] \,\,\, ,
\ee
which coincides with (\ref{Ntmax_c}) for $c=\frac{1}{2}$\,. Then we will have a positive and increasing $\ddot{a}_{(3)} (t)$ in the whole interval (\ref{NtintPio4}) for any $c < \frac{3 - \sqrt{5}}{4} \approx 0.19$\,, where the bound on $c$ is the solution to the condition $\frac{\pi}{4} < {\rm arccos} \left( \sqrt{3c-2c^2} \right)$ in the interval $0<c<\frac{1}{2}$\,.

To summarize, in order to have positive acceleration, which is the definition of inflation, we have to take $c<1$. Combined with the condition $c>0$, that was needed to ensure a real inflaton, this means that the parameter space of our class of inflationary models is:
\be \label{c_par_space}
0 < c < 1 \,\,\,\, .
\ee
In addition, for $c>\frac{1}{2}$ we have models with decreasing acceleration, while for $c<\frac{1}{2}$ we can have models, in which a period of increasing acceleration is followed by a period of decreasing one, or models with only increasing acceleration, depending on $c$ and our choice of the integration constant $N$.

As a last remark in this subsection, note that, clearly, in the new class of models we are studying one can define ``early times" as $t <\!\!< \frac{1}{N}$ and ``late times" as $t \sim \frac{1}{N}$. So, just like in the cases $(1)$ and $(2)$ considered in \cite{MSY}, it makes sense for us to investigate whether perturbations grow at late times or remain constant (or decaying).

\subsection{Inflaton potential}

Before turning to the fate of perturbations in the new class of constant-roll inflationary models, arising from cases $(3)$ and $(4)$, let us first derive the corresponding inflaton potential, for completeness. We will see that requiring the potential to be positive-definite does not impose any further restrictions on the model.

The inflaton potential $V (\phi)$ can be found from the first equation in (\ref{FriedEq}). For that purpose, we need to invert the function $\phi (t)$. From (\ref{NewModel_setup}), we find:
\be \label{tinv5}
t (\phi) = \frac{2}{N} \,{\rm arccot} \!\left( e^{\sqrt{\frac{c}{2}} \,\phi} \right) \,\,\, ,
\ee
where for convenience we have taken the plus sign as well as set $C^{\phi}_3 = 0$. Clearly, a nonzero $C^{\phi}_3$ can be restored in the final result by performing the shift $\phi$ $\rightarrow$ $\phi - C^{\phi}_3$. Similarly, the minus-sign case can be obtained by $\phi$ $\rightarrow$ $- \phi$. Now, differentiating $\phi_{(3)} (t)$ and substituting (\ref{tinv5}) in the result, we obtain:
\be \label{phi5phi5dot}
\dot{\phi} (\phi) = - \,\sqrt{\frac{2}{c}} \,\frac{N}{\sin \!\left[ 2 \,{\rm arccot} \!\left( e^{\sqrt{\frac{c}{2}} \,\phi} \right) \right]} \,\,\,\, .
\ee
This expression can be rewritten in a nicer form by using that $\sin (2 \alpha) = \frac{2 \cot \!\alpha}{ 1 + \cot^2 \!\alpha}$ for any $\alpha$, which implies that
\be
\sin \left( 2 \,{\rm arccot} \beta \right) = \frac{2 \beta}{1 + \beta^2}
\ee
for any $\beta$. Hence (\ref{phi5phi5dot}) acquires the form:
\be \label{phi5phi5dotp}
\dot{\phi} (\phi) = - \,\sqrt{\frac{2}{c}} \,N \cosh \!\left( \sqrt{\frac{c}{2}} \,\phi \right) \,\,\, .
\ee
The other ingredient we need is $H(\phi)$. Substituting (\ref{tinv5}) in $H_{(3)} (t)$, we find:
\be \label{H5phi5}
H (\phi) = \frac{N}{c} \,\cot \!\left[ 2 \,{\rm arccot} \!\left( e^{\sqrt{\frac{c}{2}} \,\phi} \right) \right] = \frac{N}{c} \,\sinh \!\left( \sqrt{\frac{c}{2}} \,\phi \right) \,\,\, ,
\ee
where in the last step we have used that 
\be
\cot \left( 2 \,{\rm arccot} \beta \right) = \frac{\beta^2 - 1}{2 \beta}
\ee
for any $\beta$. This relation can be derived from $\sin (2 \alpha) = \frac{2 \cot \!\alpha}{ \cot^2 \!\alpha + 1}$ and $\cos (2 \alpha) = \frac{\cot^2 \!\alpha - 1}{ \cot^2 \!\alpha + 1}$\,, which are true for any $\alpha$.

Finally, substituting (\ref{phi5phi5dotp}) and (\ref{H5phi5}) in the first equation of (\ref{FriedEq}), we have:
\be
V(\phi) = 3 H^2 - \frac{\dot{\phi}^2}{2} = \frac{N^2}{2 c^2} \left[ (3-c) \cosh \!\left( \sqrt{2c} \,\phi \right) - (3 + c) \right] \,\,\, .
\ee
To incorporate the nonzero arbitrary integration constant $C^{\phi}_3$, let us now shift $\phi \rightarrow \phi + \phi_0$, where we have denoted $\phi_0 \equiv - C^{\phi}_3$. The result is:
\be \label{V_sc_pot}
V(\phi) = \frac{N^2}{2 c^2} \left[ (3-c) \cosh \!\left( \sqrt{2c} \,(\phi + \phi_0) \right) - (3 + c) \right] \,\,\, .
\ee
Note that, since $0<c<1$, the first term is always positive. Then, in principle, we can ensure that $V(\phi)$ is positive-definite by choosing suitably the integration constant $\phi_0$. It turns out, however, that this is not necessary. Namely, even with $\phi_0 = 0$, the first term is always greater than the second one, within the ranges of interest for us. 

To see this, let us rewrite the potential, with $\phi_0 = 0$, as: 
\be
V (\phi) = \frac{N^2}{2 c^2} \,(3-c) \!\left[ \cosh \!\left( \sqrt{2c} \,\phi \right) - \frac{(3 + c)}{(3-c)} \right]
\ee
and compare the two terms in the bracket $V_{t1} \equiv \cosh \!\left( \sqrt{2c} \,\phi \right)$ and $V_{t2} \equiv \frac{(3+c)}{(3-c)}$. For future use, we begin by considering the interval (\ref{NtintPio4}). In this case, the minimum value of $V_{t1} = \cosh (2 \ln [\cot (\frac{N t}{2})])$ is
\be
V_{t1}|_{min} = \cosh \!\left( 2 \ln \!\left[ \cot \!\left( \frac{\pi}{8} \right) \right] \right) = 3 \,\,\, ,
\ee
which is greater than the maximum value of $\frac{(3+c)}{(3-c)}$\,, that is:
\be
V_{t2}|_{max} = \left( \frac{3+c}{3-c} \right) \!\bigg|_{c=1} \!= 2 \,\,\,\, . 
\ee
The same qualitative result is valid also when considering the more general interval (\ref{Ntmax_c}). Indeed, the minimal value of the first term now is: 
\be
V_{t1}|_{min} = \cosh \!\left( 2 \ln \!\left[ \cot \!\left( \frac{{\rm arccos} (\sqrt{c})}{2} \right) \right] \right) = \frac{1+c}{1-c} \,\,\,\, ,
\ee
which is greater than $V_{t2} = \left( \frac{3+c}{3-c} \right)$ for any $c \in (0,1)$.

To conclude, the requirement, that $V(\phi)$ is positive-definite, does not impose any further restrictions on the parameter space of our class of models.

\section{Scalar perturbations and stability} \label{SecPertSt}
\setcounter{equation}{0}

In the previous Section we showed that the new class of models, with Hubble parameter, scale factor and inflaton as in (\ref{NewModel_setup}) and potential given by (\ref{V_sc_pot}), can describe inflationary expansion. Namely, we saw that for
\be \label{ParSp_c}
0 < c < 1
\ee
one has positive acceleration in the time interval 
\be \label{posac_t_int}
t \in \left[ 0,\,\frac{1}{N} \arccos (\sqrt{c}) \!\right) \,\, . 
\ee
In fact, as explained in Subsection \ref{SecPosAccel}, we can choose any subinterval of (\ref{posac_t_int}) as part of our freedom to redefine the integration constant $N$. It will be particularly useful later on to take $t \in [0, \frac{\pi}{4N}]$, as already mentioned.

Now we want to consider the evolution of scalar perturbations in this class of models. Our purpose is twofold. First we want to show that those perturbations do not have growing modes on super-Hubble scales. This would mean that the models are stable and, therefore, the inflationary stage can last long enough to produce the necessary amount of expansion.\footnote{Recall that roughly about 60 or so e-folds are needed to account for the observed homogeneity and isotropy of the Universe.} Given that, our second goal is to derive the scalar spectral index $n_s$ for this class of models and to find a region of our parameter space, that is compatible with the value $n_s \approx 0.96$ determined from observations.

As is well-known, the scalar perturbations of the inflaton $\delta \phi$ and of the metric $\delta g_{\mu \nu}$ are not physically meaningful separately, since they are not invariant under gauge transformations that arise from certain coordinate reparametrizations. There is only one independent scalar degree of freedom, the gauge-invariant curvature perturbation $\zeta$. In comoving gauge, it relates to the above fluctuations via $\delta \phi = 0$ and $\delta g_{ij} = a^2 \left[ ( 1 - 2 \zeta) \delta_{ij} + h_{ij} \right]$, where $h_{ij}$ are the tensor perturbations and $i,j=1,2,3$; see for example \cite{DB}. It is useful to Fourier transform $\zeta$, namely: $\zeta (t,\vec{x}) = \int \!\frac{d^3 k}{(2\pi)^3} \,\zeta_k (t) \,e^{i \vec{k}.\vec{x}}$\,. Then, introducing $v_k \equiv \sqrt{2} \tilde{z} \zeta_k$ with $\tilde{z}^2 \equiv - a^2 \!\frac{\dot{H}}{H^2}$\,, the evolution of the mode function $v_k$ is determined by the Mukhanov-Sasaki equation \cite{VM,MSa}: 
\be \label{MukhSasEq}
v_k'' + \left( k^2 - \frac{\tilde{z}''}{\tilde{z}} \right) v_k = 0 \,\,\, ,
\ee
where $k \equiv |\vec{k}|$ and $'\!\equiv \!\pd_{\tau}$ with $\tau$ being conformal time. The latter is defined as usual via 
\be \label{ConfTime}
\tau = \pm \int \frac{dt}{a(t)} \,\,\, ,
\ee
to ensure that the spacetime metric becomes $ds^2_4=-dt^2+a^2d\vec{x}^2\equiv a^2\left( -d\tau^2+d\vec{x}^2 \right)$.

Finally, the term $\tilde{z}''/\tilde{z}$ in (\ref{MukhSasEq}) can be written as \cite{LL,MSY}:
\be \label{MassTerm}
\frac{\tilde{z}''}{\tilde{z}} = a^2 H^2 \left( 2 - \epsilon_1 + \frac{3}{2} \epsilon_2 + \frac{1}{4} \epsilon_2^2 - \frac{1}{2} \epsilon_1 \epsilon_2 + \frac{1}{2} \epsilon_2 \epsilon_3 \right) \,\, ,
\ee
where $\epsilon_i$ are a series of slow roll parameters defined by:
\be \label{srPar}
\epsilon_1 \equiv - \frac{\dot{H}}{H^2} \qquad {\rm and} \qquad \epsilon_{i+1} \equiv \frac{\dot{\epsilon}_i}{H \epsilon_i} \quad .
\ee
Note that the expression on the right-hand side of (\ref{MassTerm}) is exact, and not a leading contribution in the slow-roll approximation.

\subsection{Super-Hubble scales}

Let us now address the question of whether our class of models is stable or not. For that purpose, we need to investigate the behavior of the scalar perturbations on super-Hubble scales. If they were to have a growing mode at late times, that would indicate instability. We will see here, instead, that all modes are decaying or constant.

To begin, note that on super-Hubble scales one has $k^2 <\!\!< \tilde{z}''\!/\tilde{z}$ and so (\ref{MukhSasEq}) simplifies to:
\be \label{MukhSas}
v_k'' - \frac{\tilde{z}''}{\tilde{z}} \,v_k = 0 \,\,\, .
\ee
As noted in \cite{MSY}, the general solution of (\ref{MukhSas}) is such that $\zeta_k = \frac{\sqrt{2}}{2} \frac{v_k}{\tilde{z}}$ has the form:
\be \label{zeta}
\zeta_k = A_k + B_k \!\int \!\!\frac{dt}{a^3 \epsilon_1} \,\,\, ,
\ee
where $A_k,B_k=const$, regardless of what the explicit function $\tau = \tau (t)$ is. So we will investigate the behavior of (\ref{zeta}) in the following. In fact, it is convenient to rewrite the latter as:
\be \label{zetaHdot}
\zeta_k = A_k + B_k \!\int \!\!\frac{\,H^2}{a^3 \dot{H}} \, dt \quad ,
\ee
where we have used (\ref{srPar}) and, also, we have absorbed a minus sign in the arbitrary integration constant $B_k$. Now the important question is whether the magnitude of the integral in (\ref{zetaHdot}) increases or decreases at late times.

To answer this question for our class of models, let us compute the relevant integral. Substituting $H_{(3)}$ and $a_{(3)}$ from (\ref{NewModel_setup}), we obtain:
\bea \label{Icase3}
\int \!\frac{H^2}{a^3 \dot{H}} \,dt &=& - \,\frac{1}{c \,(C_3^a)^3} \int \!\frac{\cos^2 (Nt)}{\sin^{3/c} (Nt)} \,dt \\
&=& \,\frac{1}{3 \,c \,N (C_3^a)^3} \,\cos^3 (Nt) \,\, {}_2F_1 \!\left( \frac{3}{2} \,, \frac{c+3}{2c}\,, \frac{5}{2} \,; \,\cos^2 (Nt) \!\right) \,\, . \nn
\eea
Now, observe that our parameter $c$ relates to the parameter $\alpha$ in \cite{MSY} as $c = 3 + \alpha$. Then, it is immediately obvious that the indices of the hypergeometric function in (\ref{Icase3}) are exactly the same as those in case $(1)$; see eq. (47) of \cite{MSY}. Furthermore, denoting 
\be \label{xdef}
x \equiv \cos^2 (Nt) \,\,\, , 
\ee
we also see that, up to an overall numerical constant, we have in (\ref{Icase3}) the same function, namely 
\be \label{fx}
f (x) \equiv x^{\frac{3}{2}}\, {}_2F_1 \!\left( \,\frac{3}{2} \,, \frac{c+3}{2c}\,, \frac{5}{2} \,; \,x \right) \,\,\, ,
\ee
as the one in the considerations of \cite{MSY} regarding case $(1)$. 
\begin{figure}[t]
\begin{center}
\hspace*{0.1cm}\includegraphics[scale=0.395]{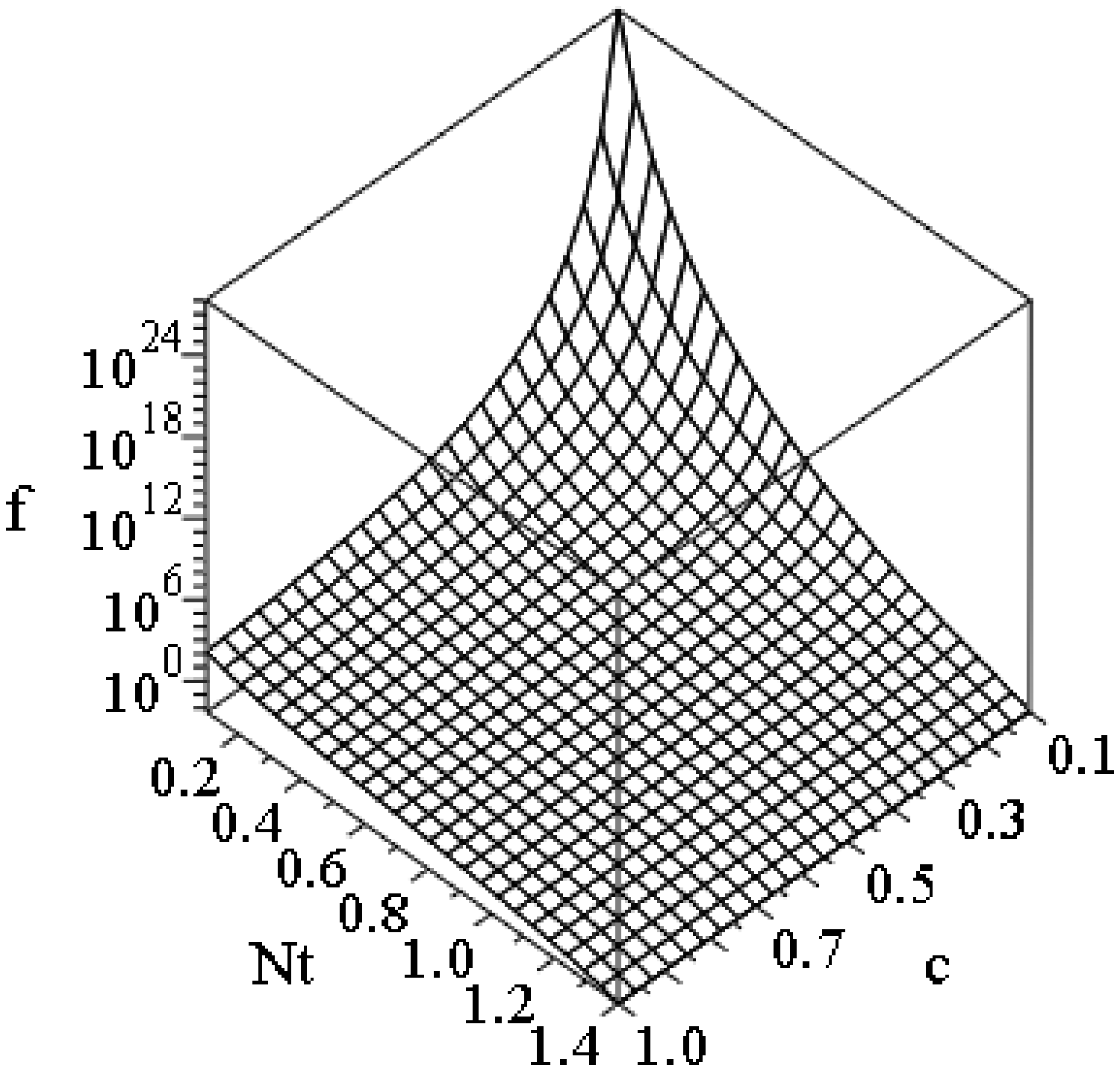}
\hspace*{-0.9cm}\includegraphics[scale=0.45]{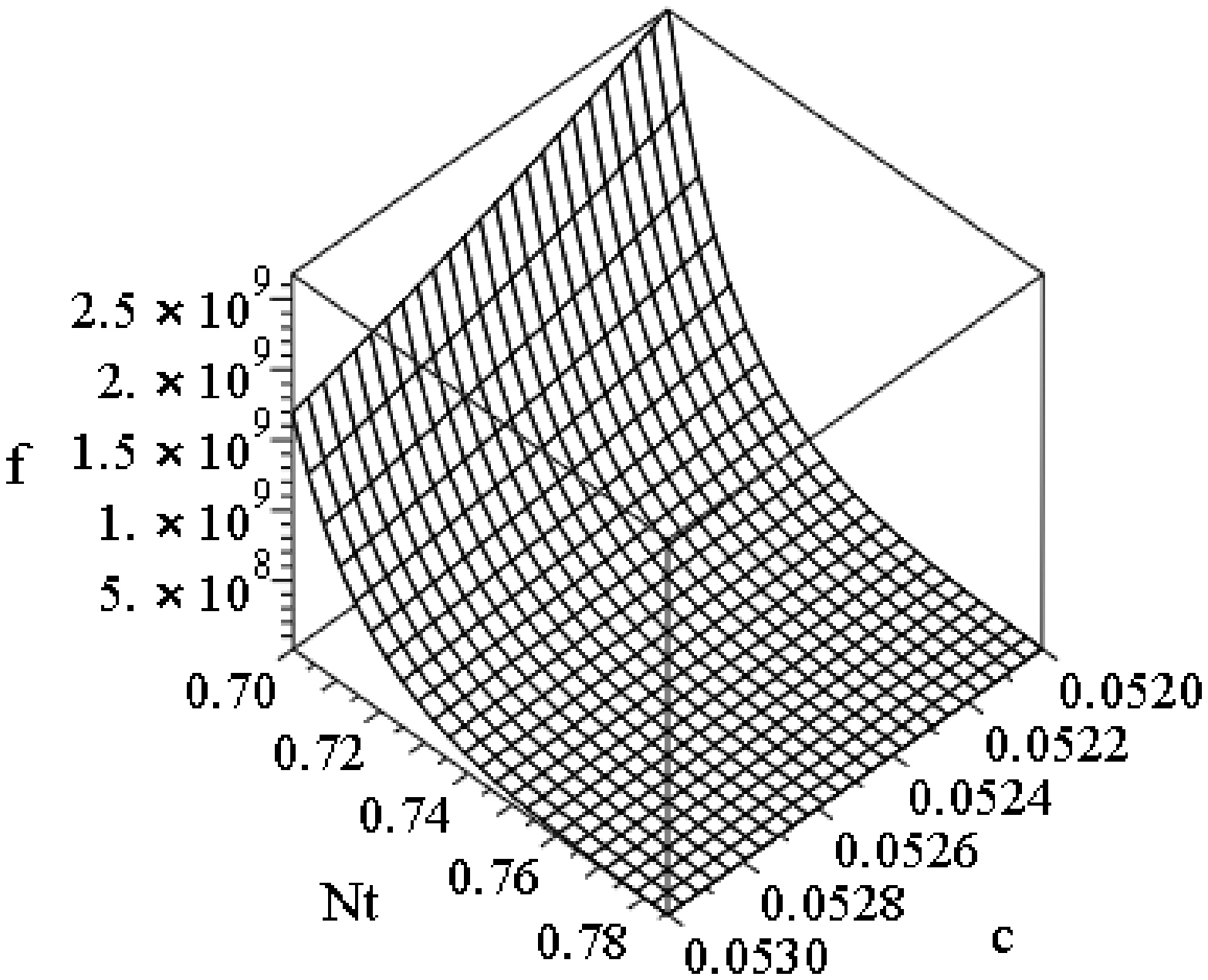}
\end{center}
\vspace{-0.6cm}
\caption{{\small Plots of $f(x)$ in (\ref{fx})-(\ref{xdef}) as a function of both the combination $Nt$ and the parameter $c$. On the left we have plotted $f(Nt,c)$ for $Nt\in (0,\frac{\pi}{2})$ and $c\in (0,1)$, while on the right we have zoomed in on a representative slice of this plot for $Nt \in [\,0.7\,,\,\frac{\pi}{4}\,]$ and $c \in [\,0.053\,,\,0.054\,]$. The particular intervals for the horizontal axes on the right will be important in Section \ref{Sec_n_s}.}}
\label{3dPlots}
\vspace{0.3cm}
\end{figure}
However, there is a crucial difference between that case and our present situation. Namely, in case $(1)$ one needed to investigate the behavior of the function $f(x)$ in (\ref{fx}) for $x \rightarrow \infty$, whereas now $x\in [c,1]$\,. The reason is that, as we saw in Section \ref{SecPosAccel}, to ensure $\ddot{a} (t)>0$ we have to consider the interval $Nt \in [0, \arccos (\sqrt{c})]$ with $c \in (0,1)$\,. Given the definition (\ref{xdef}), obviously $x$ {\it decreases} monotonically as $t$ {\it increases}, with $x=1$ corresponding to $t=0$. So large $t$, or equivalently late times, corresponds to a neighborhood of $x=c$, which is a completely regular point for the function ${}_2F_1 \!\left( \,\frac{3}{2} \,, \frac{c+3}{2c}\,, \frac{5}{2} \,; \,x \right)$\,. Note also that, for $c \rightarrow 0$, the interval $Nt \in [0, \arccos (\sqrt{c})]$ tends to $Nt \in [0,\frac{\pi}{2})$, i.e. the maximal interval in which $H_{(3)} (t) > 0$\,.

In Section \ref{SecPosAccel} we explained that choosing the upper bound of the $t$-interval to be  $\theta_*/N$ with some $\theta_* < \arccos (\sqrt{c})$ is part of our freedom to rescale the integration constant $N$. So, just like taking a particular value of $c$, choosing a particular subinterval $Nt \in [0,\theta_*]$ is part of specifying precisely which model, within our class of models, one is considering. Nevertheless, now we will see that the perturbation modes in (\ref{Icase3}) decay for any $c$ and any subinterval, within our parameter space. In fact, the function determining their behavior, namely $f(x)$ in (\ref{fx}) with $x$ as in (\ref{xdef}), is a decreasing function of $t$ in the entire maximal interval $Nt \in [0,\frac{\pi}{2}]$ for any $0<c<1$.

\begin{figure}[t]
\begin{center}
\includegraphics[scale=0.31]{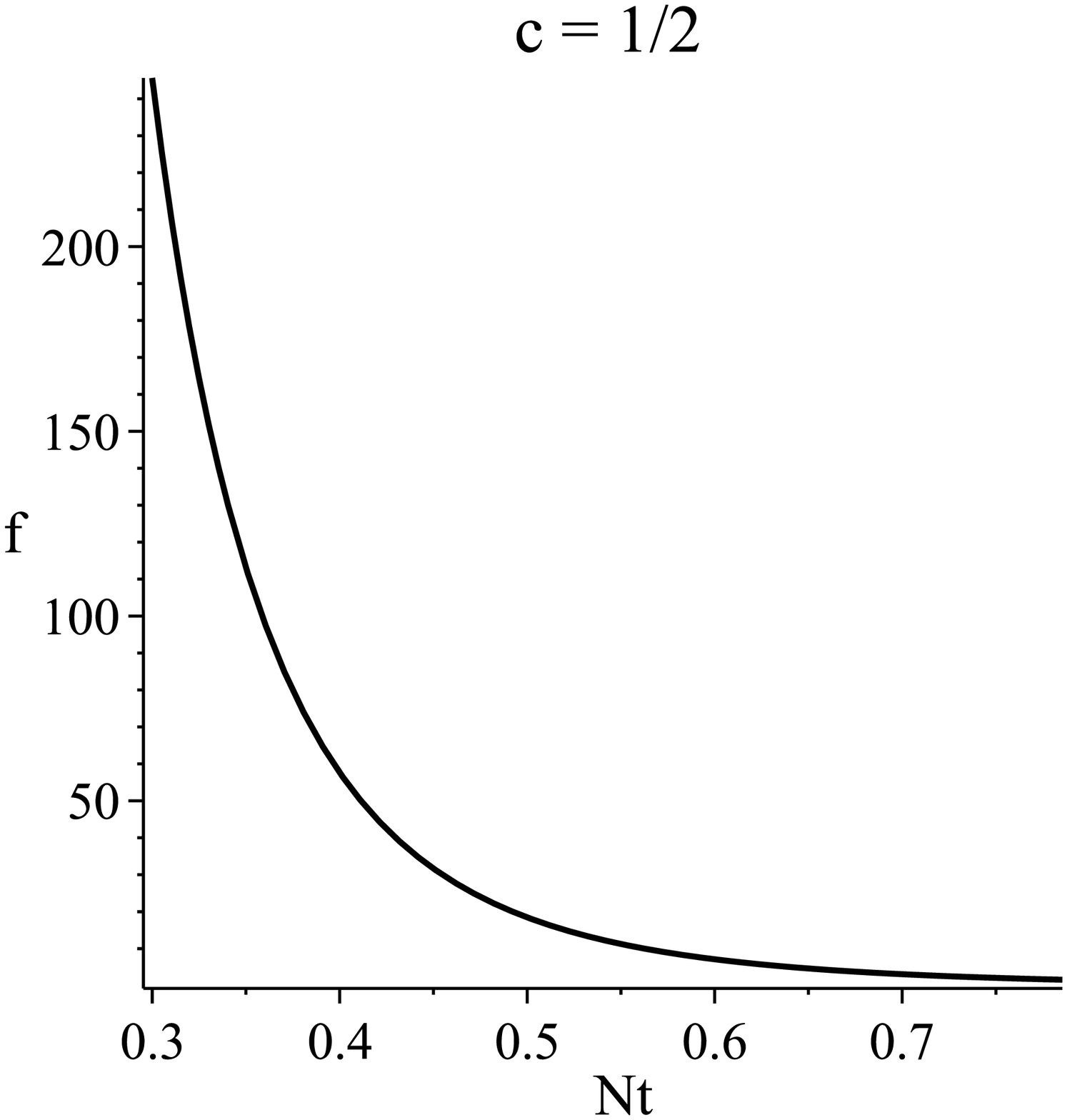}
\hspace*{0.5cm}\includegraphics[scale=0.33]{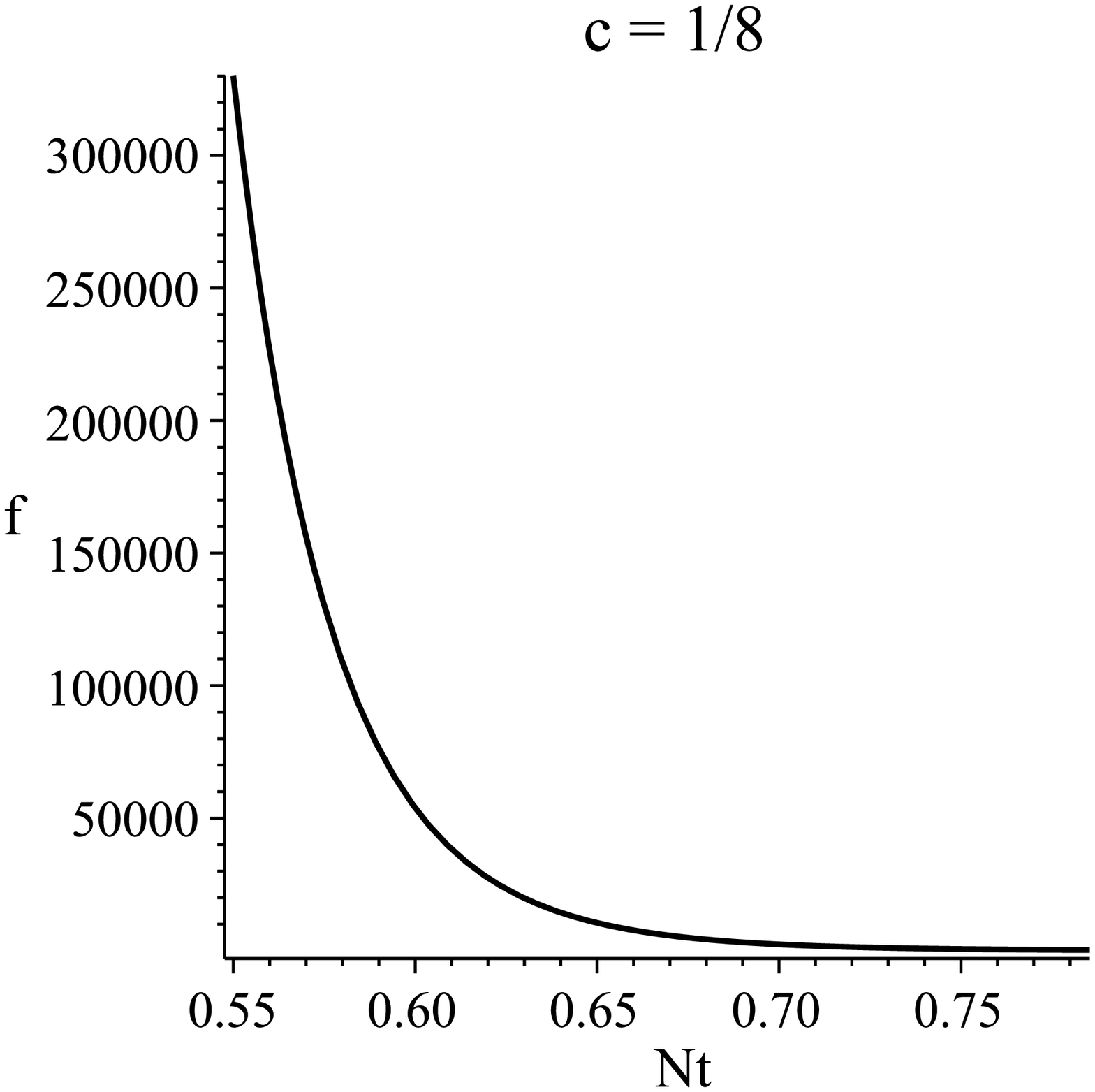}
\end{center}
\vspace{-0.6cm}
\caption{{\small Two plots of the function $f(Nt)$ for fixed values of the parameter $c$. On the left: $c=\frac{1}{2}$ and $Nt \in [\,0.3\,,\,\frac{\pi}{4}\,]$. On the right: $c=\frac{1}{8}$ and $Nt \in [\,0.55\,,\,\frac{\pi}{4}\,]$.}}
\label{f_fixed_c}
\vspace{0.2cm}
\end{figure}

One can easily verify the last statement in the following way. First, note that $f(t)$ in (\ref{fx})-(\ref{xdef}) depends only on the combination $Nt$ and the parameter $c$. Now, both $Nt$ and $c$ vary in finite intervals. Therefore, one can just plot $f$ as a function of $Nt$ {\it and} $c$, thus explicitly showing that it decreases with time for any $c\in (0,1)$, regardless of the choice of value for the constant $N$. On the left side of Figure \ref{3dPlots}, we have plotted the behavior of the function $f(Nt,c)$ for the full ranges $Nt\in (0,\frac{\pi}{2})$ and $c\in(0,1)$, using a logarithmic vertical axis since $f$ diverges very fast for either $Nt \rightarrow 0$ or $c\rightarrow 0$. On the right side, we show a representative slice of that plot, for ranges on the horizontal axes that will be useful in the next subsection. One can easily check that the behavior of $f(Nt,c)$ is qualitatively the same, as in the right plot of Figure \ref{3dPlots}, for any other subintervals of $Nt\in [0,\frac{\pi}{2}]$ and $c\in[0,1]$. In Figure \ref{f_fixed_c}, we also illustrate the decrease of $f$ with time for two fixed values of $c$. 
Again, one can easily check that, for any other choice of $c$, the functional dependence of $f(Nt)$ is qualitatively the same as in those two plots. Note that for smaller $c$ the function $f$ has greater values at early times, i.e. for $t <\!\!< \frac{1}{N}$\,, and decreases more precipitously with $t$. Also, at the upper end of the maximal $Nt$-interval we have $f|_{Nt = \frac{\pi}{2}} = 0$\,, since $x = 0$ and \,${}_2F_1 \!\left( \frac{3}{2}, \frac{c+3}{2c}, \frac{5}{2}; 0 \right) = 1$ \,for $\forall c$\,, and thus the perturbation modes completely vanish there. For any model, with $Nt$ running in a subinterval $[0,\theta_*]$ with some $\theta_* \!< \frac{\pi}{2}$\,, it is also not a problem to have as small a magnitude of the perturbations at late times (i.e., for $t\sim \frac{1}{N}$) as desired, since we have the arbitrary integration constant $C_3^a$ in the denominator of (\ref{Icase3}). So by choosing appropriately the value of $C_3^a$, we can ensure that the decreasing perturbation modes in (\ref{Icase3}) become sufficiently small in a neighborhood of $Nt = \theta_*$ for any $\theta_* \!< \frac{\pi}{2}$\,.

\begin{figure}[t]
\begin{center}
 \includegraphics[scale=0.65]{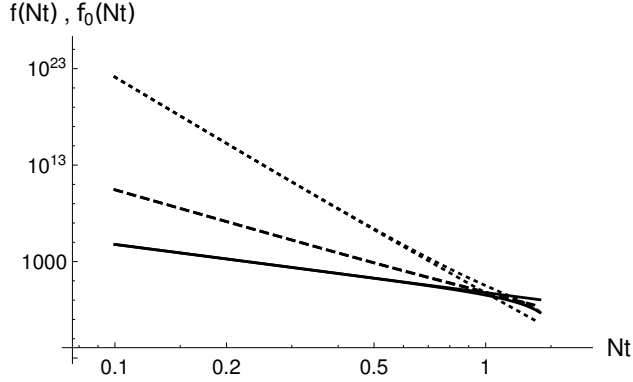}
 \caption{{\small Logarithmic plot of $f(Nt)$ and its approximation $f_0(Nt)$ in the interval $Nt \in (0,\frac{\pi}{2})$ for three values of the parameter $c$\,: for $c=\frac{1}{2}$ (solid line), $c=\frac{1}{4}$ (dashed line) and $c=\frac{1}{8}$ (dotted line); the exact curve is the lower one, when they can be distinguished.}}
 \label{ff_0}
\end{center}
\end{figure}
So far, we have verified numerically that the perturbation modes in (\ref{Icase3}) are decaying with time for the entire parameter space of our class of models. To gain some analytical insight into the reason for this, let us consider the function $f \left( x(Nt) \right)$ in (\ref{fx})-(\ref{xdef}) in a neighborhood of its singular point $Nt = 0$. It has the following expansion:
\be \label{FNt_expan}
f(Nt) = (Nt)^{1-\frac{3}{c}}\!\left[\frac{3\,c}{3-c}+\frac{1-2\,c}{2\,(1-c)}(Nt)^2 + \frac{(15 - 58 c + 40 c^2)}{40\,c\,(3 - 5 c)} (Nt)^4 + ...\right] + const \, ,
\ee
where the expression in the brackets is a regular function for $Nt \approx 0$\,. Now, one can verify numerically that taking the leading two terms of that Taylor expansion, namely:
\be
f_0 (Nt) = \frac{3\,c}{3-c} \,(Nt)^{1-\frac{3}{c}} + \frac{1-2\,c}{2\,(1-c)} \,(Nt)^{3-\frac{3}{c}} \,\,\,\, ,
\ee
gives a rather good approximation for $f(Nt)$ in the whole interval $Nt \in (0,\frac{\pi}{2}]$ when $c\le \frac{1}{2}$. To illustrate that point economically, we have plotted in Figure \ref{ff_0} both $f(Nt)$ and $f_0 (Nt)$ on a logarithmic scale for three different values of the parameter $c$: for $c=\frac{1}{2}$ (solid line), $c=\frac{1}{4}$ (dashed line) and $c=\frac{1}{8}$ (dotted line). Note that, to obtain a good approximation for $c>\frac{1}{2}$, one needs to take also the third term in (\ref{FNt_expan}). From the above considerations, we can conclude that the behavior of $f(t)$ in the interval $t\in [0,\frac{\pi}{2N}]$ is dominated by the singularity for $t \rightarrow 0$, at which $f (t) \rightarrow +\infty$\,. Combined with the vanishing of $f (t)$ at $t = \frac{\pi}{2N}$ and the fact that $f(t)$ is regular at any $t \in (0,\frac{\pi}{2N}]$, this leads to rapid decrease with time throughout the entire interval.

To summarize, we have seen that the perturbations in our new class of inflationary models either remain constant, which is represented by the $A_k=const$ term in (\ref{zetaHdot}), or rapidly decay with time, for any $0\!<\!c\!<\!1$ and any interval $t\in [0,\frac{\theta_*}{N}]$ with $\theta_* \!\le \frac{\pi}{2}$\,. Thus, these models are stable. Finally, let us underline again that, by suitably choosing the integration constant $N<\!\!<1$, we can ensure that the duration of the inflationary stage is as large as desired.

\subsection{Scalar spectral index} \label{Sec_n_s}

Now that we have established that the class of models with (\ref{NewModel_setup}), (\ref{V_sc_pot}) and (\ref{ParSp_c})-(\ref{posac_t_int}) describes stable inflationary expansion, our goal will be to find a part of its parameter space, in which it can give a spectral index $n_s$ compatible with observations. Recall that choosing a specific model within this class amounts both to picking a value for $c$ as well as taking a particular subinterval of (\ref{posac_t_int}).

To compute $n_s$\,, one needs to study the perturbation equation (\ref{MukhSasEq}) on scales, on which both the $k^2$ and the $\frac{\tilde{z}''}{\tilde{z}}$ terms are not negligible. For non-slow roll inflationary models, this can be very difficult generically, due to the dependence of $\tilde{z}$ on the background. So, in principle, it would require numerical investigation \cite{DB}. However, we will show that, for a certain subclass of models within our class, one can obtain an analytical estimate. In addition, we will then argue that this result is likely to provide, actually, the best fit to observations for the entire class of models.

To be more specific, in the following we will consider models with small $c$\,, meaning at least $c<\frac{1}{2}$\,, and also we will take the interval:
\be \label{t_int_Pi_o_4}
t \in \left[ 0, \frac{\pi}{4N}\right] \,\,\, .
\ee
According to the discussion in Section \ref{SecPosAccel}, this means that the acceleration $\ddot{a}(t)$ can be increasing in a part (or all) of the interval. In fact, as pointed out below equation (\ref{NtintPio4}), for $c \lesssim 0.19$ \,the acceleration increases in the whole interval (\ref{t_int_Pi_o_4}), similarly to the familiar de Sitter case.

Let us now compute the different ingredients we need to study equation (\ref{MukhSasEq}). We begin by finding conformal time $\tau$. Substituting $a_{(3)}$ from (\ref{NewModel_setup}) in (\ref{ConfTime}), we obtain:
\be
\int \!\frac{dt}{a_{(3)}} \,= \,- \,\frac{1}{C_3^a N} \,\,\cos (Nt) \,\,{}_2 F_1 \!\left( \frac{1}{2}\,, \,\frac{c+1}{2c}\,, \,\frac{3}{2}\,; \,\cos^2 (Nt) \!\right) \,+ \,\,const \,\,\, .
\ee
This expression tends to $-\infty$ as $t\rightarrow 0$ and is a negative constant at $t_{max} = \frac{\pi}{4N}$. So it is convenient to choose the integration constant such that
\be \label{tau_function}
\tau \,= \,- \,\frac{1}{C_3^a N} \left[ \cos (Nt) \,\,{}_2 F_1\!\left( \frac{1}{2}\,, \,\frac{c+1}{2c}\,, \,\frac{3}{2}\,; \,\cos^2 (Nt) \!\right) - \frac{\sqrt{2}}{2} \,\,{}_2 F_1\!\left( \frac{1}{2}\,, \,\frac{c+1}{2c}\,, \,\frac{3}{2}\,; \,\frac{1}{2} \right) \right] \,\, .
\ee
This way we have:
\be \label{tau_range}
\tau \in (-\infty \,, \,0\,]
\ee
as $t$ runs in the interval in (\ref{t_int_Pi_o_4}). Note that this range for $\tau$ is the same as in cases $(1)$ and $(2)$; see \cite{MSY}. It is also worth pointing out that this is the same range as for the conformal time in pure de Sitter inflation (i.e. when $H = const$).

As a last remark on the form of the function $\tau = \tau (t)$, note that if we were considering the maximal interval (\ref{posac_t_int}), in which $\ddot{a} (t) > 0$, then we could take the integration constant such that
\be \label{tau_const_c}
\tau \,= \,- \,\frac{1}{C_3^a N} \left[ \,\cos (Nt) \,\,{}_2 F_1\!\left( \frac{1}{2}\,, \,\frac{c+1}{2c}\,, \,\frac{3}{2}\,; \,\cos^2 (Nt) \!\right) \,- \,\,\sqrt{c} \,\,\,{}_2 F_1\!\left( \frac{1}{2}\,, \,\frac{c+1}{2c}\,, \,\frac{3}{2}\,; \,c \right) \right]\,\, ,
\ee
ensuring again that $\tau$ runs in (\ref{tau_range}); obviously, the same logic goes for any other subinterval of (\ref{posac_t_int}). So this nice range for $\tau$ is not a special artifact of the choice (\ref{t_int_Pi_o_4}), but a generic feature of the entire class of models.

Now, let us also find the slow-roll parameters $\epsilon_{1,2,3}$ that enter equation (\ref{MassTerm}). Substituting $H_{(3)}$ form (\ref{NewModel_setup}) in (\ref{srPar}), we obtain:
\be \label{epsilon_i_exact}
\epsilon_1 = \frac{c}{\cos^2 (Nt)} \quad , \quad \epsilon_2 = 2 c \,\tan^2 (Nt) \quad , \quad \epsilon_3 = \frac{2c}{\cos^2 (Nt)} \quad .
\ee
It is also easy to realize that:
\be
\epsilon_{2j} = \epsilon_2 \qquad \,\,{\rm and} \,\,\qquad \epsilon_{2j+1} = \epsilon_3 \qquad \,\,{\rm for} \quad\forall j \ge 1 \qquad .
\ee
Clearly, all slow-roll parameters are non-vanishing. However, note that, for $t$ as in (\ref{t_int_Pi_o_4}), the $\epsilon_i$'s vary within the following ranges:
\be \label{epsilon_i_ranges}
c \le \epsilon_1 \le 2 c \quad \,\,, \quad \,\,0 \le \epsilon_2 \le 2 c \quad \,\,, \quad \,\,2 c \le \epsilon_3 \le 4 c \quad .
\ee
Obviously, the ranges in (\ref{epsilon_i_ranges}) become quite short numerically, if one considers models with
\be \label{small_c}
c<\!\!<1 \,\,\, .
\ee
In that case, then, one can view the parameters $\epsilon_i$ as approximately constant during the entire inflationary stage. Clearly, the smaller $c$ is, the better this approximation is. We will see at the end that our result, for values of $c$ that give $n_s\approx 0.96$\,, will be consistent with this approximation.

We are finally ready to turn to solving equation (\ref{MukhSasEq}). Let us take, as usual, the initial condition that
\be \label{initial_cond}
v_k (\tau) = \frac{e^{-i k \tau}}{\sqrt{2 k}} \qquad {\rm for} \qquad \tau \rightarrow -\infty \quad .
\ee
This obviously satisfies the perturbation equation at early times, when the mode with a wave number $k$ is still well within the Hubble radius (i.e. when $k^2 >\!\!> \frac{\tilde{z}''}{\tilde{z}}$) and thus $\frac{\tilde{z}''}{\tilde{z}}$ is negligible. To find the spectral index, we need the solution at later times, around the time of horizon crossing when neither of $k^2$ and $\frac{\tilde{z}''}{\tilde{z}}$ in (\ref{MukhSasEq}) can be neglected. For that purpose, let us see how the right hand side of (\ref{MassTerm}) simplifies within the approximation (\ref{small_c}). As we explained above, in that case the slow-roll parameters are nearly constant throughout the whole time-interval under consideration. In view of \cite{WK}, we will then approximate $\epsilon_i$ in (\ref{epsilon_i_exact}) with their values in the upper half of the interval:
\be \label{epsilon_i_approx}
\epsilon_1 \approx 2 c \quad \,\,, \quad \,\, \epsilon_2 \approx 2 c \quad \,\,, \quad \,\, \epsilon_3 \approx 4 c \quad .
\ee
Then, in (\ref{MassTerm}) we have the expression:
\be \label{Expr_epsilons}
\left( 2 - \epsilon_1 + \frac{3}{2} \epsilon_2 + \frac{1}{4} \epsilon_2^2 - \frac{1}{2} \epsilon_1 \epsilon_2 + \frac{1}{2} \epsilon_2 \epsilon_3 \right) \approx \,2 + c + 3 c^2 = \left( \frac{9}{4} + c + 3 c^2 \right) - \frac{1}{4} \,\,\,\, ,
\ee
where the rewriting in the last step, similar to \cite{MSY}, is for future convenience. Another important consequence of (\ref{small_c}) is that 
\be \label{tau_approx_aH}
-\tau \approx \frac{1}{a_{(3)} H_{(3)}} \,\,\, . 
\ee
Note that, in the de Sitter case (i.e. with $H = const$), one has exactly $\tau = - \frac{1}{aH}$\,. 
\begin{figure}[t]
\begin{center}
\hspace*{-0.1cm}\includegraphics[scale=0.38]{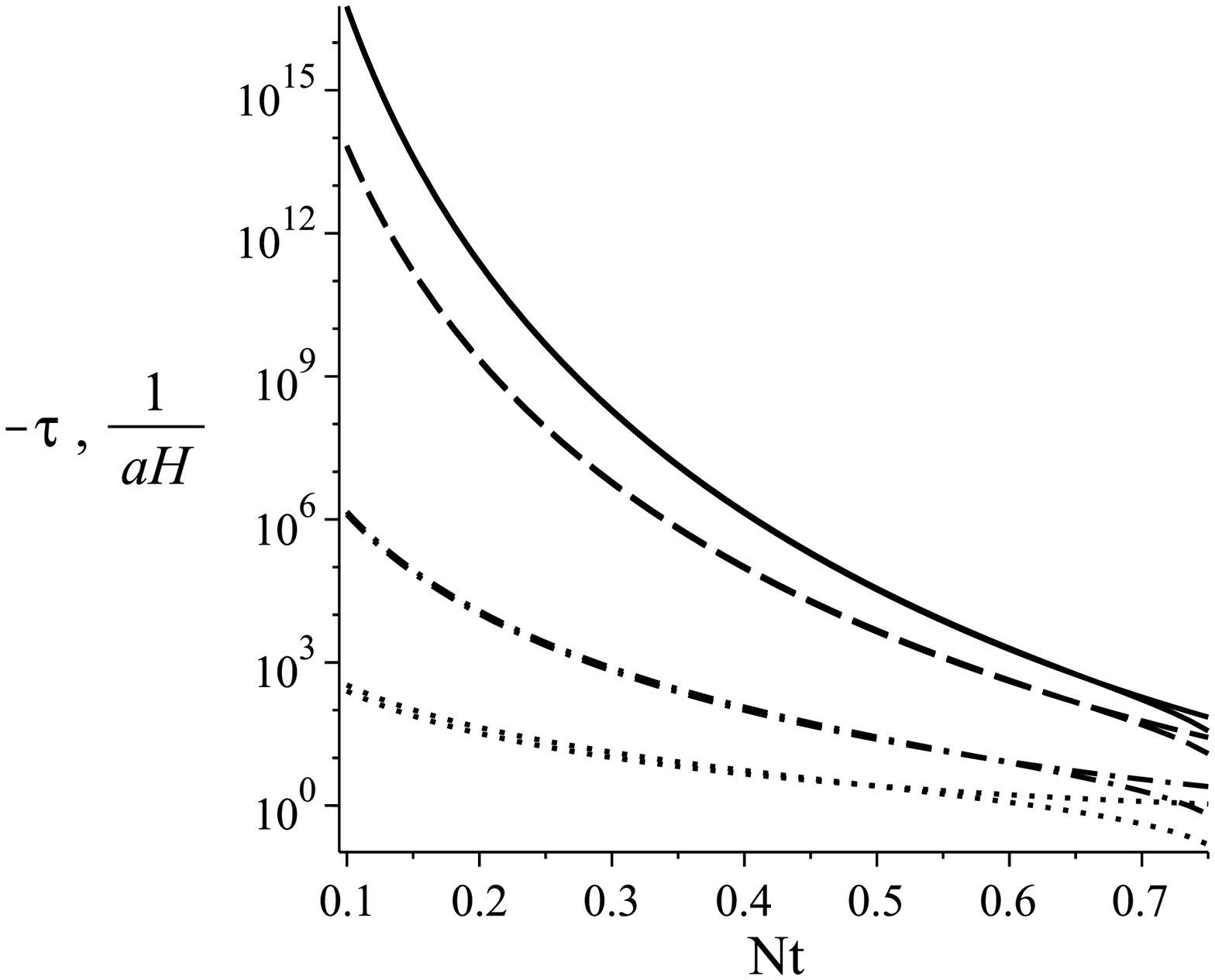}
\hspace*{0.1cm}\includegraphics[scale=0.38]{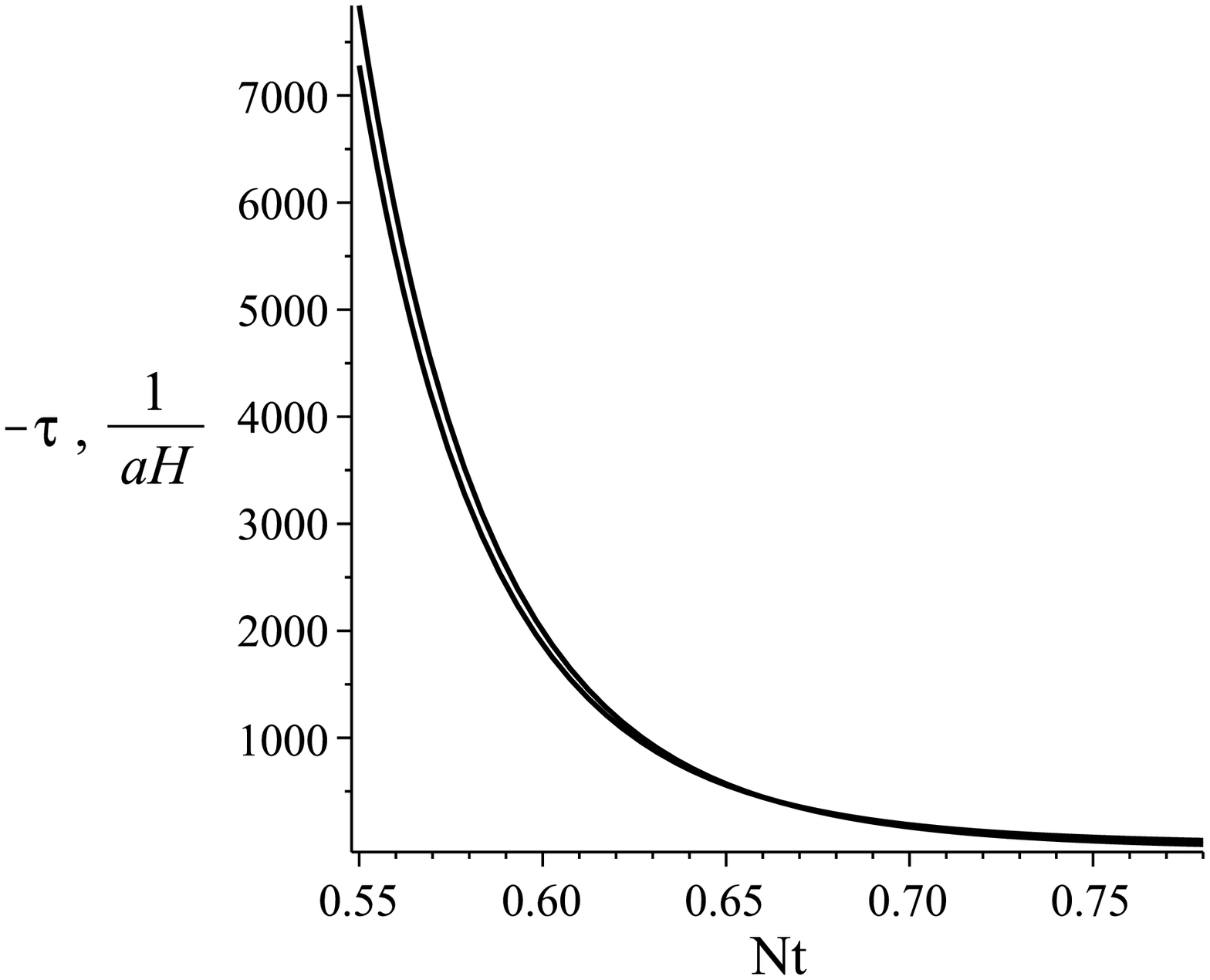}
\end{center}
\vspace{-0.6cm}
\caption{{\small Plots of $-\tau$ and $\frac{1}{aH}$ in the interval $Nt \!\in \!(0,\frac{\pi}{4})$ for several values of $c$\,, with an overall multiplier \!{\footnotesize $(C_3^a N)^{-1}$} \!factored out of both functions. On the left, the vertical axis is logarithmic and the four pairs of curves correspond to: $c=\frac{1}{4}$ (dotted), $c=\frac{1}{8}$ (dotted-dashed), $c=\frac{1}{16}$ (dashed) and $c=\frac{1}{19}$ (solid), with $-\tau$ being the lower (at late times) curve in each pair. Obviously, the approximation (\ref{tau_approx_aH}) improves with decreasing $c$. On the right $c=\frac{1}{19}$ and $-\tau$ is the higher (at early times) curve. The horizontal axis range is shortened since, otherwise, the two curves are not distinguishable at all.}}
\label{tau_vs_aH}
\vspace{0.2cm}
\end{figure}
For non-slow roll inflation this does not need to be the case. However, for our class of models, one can verify numerically that, when $c$ is small, (\ref{tau_approx_aH}) is a good approximation in the entire interval (\ref{t_int_Pi_o_4}). And the approximation is improving with the decrease of $c$. We have illustrated that on Figure \ref{tau_vs_aH} for several choices of $c$ .\footnote{Note that both the conformal time $\tau$ in (\ref{tau_function}) and the comoving Hubble radius $(a_{(3)}H_{(3)})^{-1}$ from (\ref{NewModel_setup}) have the same overall multiplier $(C_3^a N)^{-1}$, which is thus irrelevant for the comparison between the two. Factoring it out, one is left with functions of {\it only} the combination $Nt$ and the parameter $c$. These are the functions we have plotted in Figure \ref{tau_vs_aH}.} Taking into account (\ref{Expr_epsilons}) and (\ref{tau_approx_aH}), we find from (\ref{MassTerm}):
\be \label{z_simpl_nu_def}
\frac{\tilde{z}''}{\tilde{z}} \approx \frac{\nu^2 - \frac{1}{4}}{\tau^2} \quad , \qquad {\rm where} \quad \,\,\nu^2 \equiv \frac{9}{4} + c + 3 c^2 \,\,\,\, .
\ee
Then, the solution of (\ref{MukhSasEq}) with the initial condition (\ref{initial_cond}) is given by the standard expression:
\be
v_k (\tau) = \frac{\sqrt{\pi}}{2} \,\sqrt{-\tau} \,\,H^{(1)}_{\nu} (-k \tau) \,\,\,\, ,
\ee
where $H^{(1)}_{\nu}$ is the Hankel function of the first kind. Hence, the usual computation of the scalar power spectrum $\Delta_s^2 (k) \sim k^{n_s -1}$\,, reviewed in \cite{MSY}, gives for the spectral index the standard relation $n_s - 1 = 3 - 2\nu$. Substituting $\nu$ from (\ref{z_simpl_nu_def}), we then find:
\be \label{n_s_expr}
n_s = 4 - 2 \sqrt{\frac{9}{4} + c + 3 c^2} \,\,\,\, .
\ee
Recall that $c$ is small here. Thus, our expression for $\nu$ in (\ref{z_simpl_nu_def}) represents a small deviation around the de Sitter result $\nu = \frac{3}{2}$\,. Now, requiring that $n_s = 0.96$\,, the value extracted from observational data, we obtain:
\be \label{c_obs}
c = 0.0522 \approx \frac{1}{19} \,\,\,\, .
\ee
We have discarded the other solution of the quadratic equation, arising from (\ref{n_s_expr}), since it gives $c<0$.

Finally, let us comment on whether the result (\ref{c_obs}) is consistent with the approximations we used in order to obtain it. One can easily see that, for $c\approx \frac{1}{19}$\,, the slow roll parameters vary, throughout the entire interval (\ref{t_int_Pi_o_4}), within the ranges:
\be
0.05 \lesssim \epsilon_1 \lesssim 0.1 \quad \,\,, \quad \,\,0 \le \epsilon_2 \lesssim 0.1 \quad \,\,, \quad \,\,0.1 \lesssim \epsilon_3 \lesssim 0.2 \quad .
\ee
So, indeed, the approximation in (\ref{epsilon_i_approx}) is very good. In addition, one can see on Figure \ref{tau_vs_aH}, that the approximation in (\ref{tau_approx_aH}) is also satisfied to an excellent degree. Hence, we have shown that, for $c \approx \frac{1}{19}$\,, the new class of constant-roll inflationary models is consistent with observations.\footnote{Obviously, since the observational value $n_s = 0.96$ is determined only up to some precision, the requirement for consistency with observations does not pick out a single value of the parameter $c$\,, but instead a corresponding (short) interval around $c=0.0522$\,\,.} 

Note that (\ref{n_s_expr}) implies that greater values of $c$ give smaller $n_s$\,. This seems to suggest that considering \,$\frac{1}{2} \!\lesssim c <\!1$ \,is unlikely to give a result consistent with the observational constraints. Of course, the formula (\ref{n_s_expr}) was derived for small $c$, which implied that (\ref{tau_approx_aH}) and $\epsilon_i \approx const$ were good approximations. For $c\gtrsim \frac{1}{2}$\,, those approximations are not particularly good. So one would have to solve the Mukhanov-Sasaki equation numerically, in order to make definitive conclusions about greater values of $c$. We hope to come back to this investigation in the future.

\section{Discussion} \label{SecDisc}
\setcounter{equation}{0}

We investigated systematically the constant-roll condition (\ref{Heq}), obtaining in the process a solution that was missed in \cite{MSY}. We also showed that the four real solutions in (\ref{Hsol}) can be written in a unified manner as (\ref{HNewSol}), with complex integration constants. At first sight, the unifying expression may seem to contain more information than (\ref{Hsol}). However, we proved in Appendix \ref{HComplexConsts} that (\ref{HNewSol}) is real only for very specific choices of the two integration constants. Those special choices give only the solutions (\ref{Hsol}), in accordance with (\ref{hk4cases}). Thus, restricted to real values, the general expression is equivalent to the set of four real solutions. Having the unifying form of the Hubble parameter (\ref{HNewSol}), however, is rather useful when considering the resulting inflaton solution, as we saw in Subsection \ref{SecInflaton} and Appendix \ref{ComplexArctanh}.

The first two cases in (\ref{Hsol}) were investigated in detail in \cite{MSY}, which showed that in case (2) one can find parameter values such that the inflationary model is phenomenologically viable. Here we focused on cases $(3)$ and $(4)$. We saw that, to have a real inflaton, one needs $c>0$, where $c$ is the parameter in (\ref{Heq}). Then, the physical requirement for a positive Hubble parameter $H(t)$ implies that, in fact, cases $(3)$ and $(4)$ give rise to the same class of models. The functions characterizing these models are (\ref{NewModel_setup}):
\be
H (t) = \frac{N}{c} \cot (Nt) \,\,\,\, , \,\,\,\, a (t) = C_a \sin^{1/c} (N t) \,\,\,\, , \,\,\,\, \phi (t) = \pm \sqrt{\frac{2}{c}} \ln \!\left[ \cot \!\left( \frac{Nt}{2} \right) \!\right] + C_{\phi} \,\,\, ,
\ee
as well as the scalar potential (\ref{V_sc_pot}):
\be
V(\phi) = \frac{N^2}{2 c^2} \left[ (3-c) \cosh \!\left( \sqrt{2c} \,(\phi - C_{\phi}) \right) - (3 + c) \right] \,\,\, .
\ee
We showed that the acceleration $\ddot{a} (t)$ is positive in the $Nt$-interval (\ref{Ntmax_c}), or equivalently the time interval (\ref{posac_t_int}) ,\footnote{Obviously, by taking the integration constant $N<\!\!<1$\,, we can ensure that the time-interval in (\ref{posac_t_int}) is as large as desired.} when the physical parameter $c<1$\,. $\!\!$Thus, the inflationary parameter space is (\ref{c_par_space}):
\be \label{c_0_1}
0 < c < 1 \,\,\,\, .
\ee
We also pointed out that taking a subinterval of (\ref{Ntmax_c}) amounts just to rescaling the integration constant $N$. Hence, choosing a subinterval of the maximal $Nt$-interval, just as taking a concrete value of $c$, is part of specifying which particular model within this class one is considering. In addition, for all of those subintervals the duration of the inflationary period, in terms of conformal time $\tau$, can be set to:
\be
\tau \in (-\infty \,, \,0\,] \,\,\, ,
\ee
just as in the familiar de Sitter case; see (\ref{tau_range})-(\ref{tau_const_c}) and the discussion there.

Furthermore, we investigated the behavior of scalar perturbations in this new class of constant-roll inflationary models. We showed that on super-Hubble scales there are no growing modes at late times. Thus, the constant roll regime is stable and so can give the necessary amount of e-folds of expansion. This conclusion was reached for any value of $c$ in (\ref{c_0_1}) and for any subinterval of (\ref{Ntmax_c}). Then, we found a part of the parameter space of these models, in which one can obtain a scalar spectral index $n_s \approx 1$\,, as required for agreement with observations. This part is characterized by the subinterval (\ref{NtintPio4}), or equivalently the time interval (\ref{t_int_Pi_o_4}). The latter choice was convenient for enabling us to find an analytical estimate (\ref{n_s_expr}) for the scalar spectral index. Requiring $n_s = 0.96$\,, then gives $c\approx \frac{1}{19}$ within the parameter space (\ref{c_0_1}); see (\ref{c_obs}). This result was obtained within a small $c$ \,approximation that allowed us to treat the slow roll parameters $\epsilon_i$ in (\ref{epsilon_i_exact}) as constants. Note that all of our $\epsilon_i$ are non-vanishing, and not $<\!\!<1$ either, albeit for $c \approx \frac{1}{19}$ they are $\epsilon_i < 1$ for $\forall i$ during the entire inflationary period. Thus, in a sense, this is an even greater deviation from slow roll than the models considered in \cite{MSY}, for which the odd-order slow roll parameters were vanishing during the generation of the curvature perturbations. Also, our value $c=0.052$ seems closer to being genuinely intermediate, between the pure ultra-slow roll case ($c=3$) and standard slow roll ($c<\!\!<1$), than the result in \cite{MSY,MS}, which is quite close to (albeit distinct from) slow roll. Thus, the new class of models studied here is even more appealing for exploring non-slow roll inflation.

It should be pointed out that it is worth investigating the entire parameter space of these models, i.e. with $c$ not small and in particular with $c\gtrsim \frac{1}{2}$\,, as well as other subintervals of (\ref{Ntmax_c}). Most likely, this will require numerical investigation of the mode equation (\ref{MukhSasEq}). So we leave it for the future. Note, however, that even if one does not find another region of parameter space, which can give $n_s\approx 1$, it may still be useful to consider applications in the context of the low multiple-moment anomaly of the CMB, as mentioned in the introduction. It is also interesting to understand how the constant roll regime fits into the framework of the Effective Field Theory of Inflation of \cite{EFTofI} {}\footnote{For an interesting investigation along those lines, in a class of modified theories of gravity, see \cite{MH}.} and, in particular, whether the models we studied here can give appreciable modifications to the speed of sound of the scalar perturbations, in some regions of parameter space. 

Another interesting question is whether one can obtain a constant roll regime in composite inflation models \cite{CJS,BCJS}.\footnote{Studying those is timely in view of \cite{ChK},
as well as the multitude of relevant current observations.} 
Note that an ultra slow roll solution was already found in \cite{LA} with the methods of the gauge/gravity duality, applied to building a gravity dual of glueball inflation \cite{BCJS}. Since ultra-slow roll is a special (albeit unstable) example of constant roll, it makes sense to expect that constant-roll models of this type should exist as well.\footnote{Although, 
finding gravity duals in this set up, established in \cite{ASW}, has turned out to be rather nontrivial. In particular, it is not yet known what is the gravity dual of regular slow-roll composite models, despite some recent progress in that direction \cite{LA2}; see also the discussion at the end of \cite{LA3}.} Finally, more remains to be done in comparing our new class of inflationary models to observations, especially along the lines of \cite{MS} regarding the tensor-to-scalar ratio relevant for primordial gravity waves. We hope to address those issues in the near future.

\section*{Acknowledgements}

We would like to thank A. Buchel, C. Burgess, L. McAllister and D. Minic for interesting and useful conversations on various aspects of inflation. L.A. is grateful to the Simons Workshop in Mathematics and Physics, Stony Brook 2017, for hospitality during work on this project. L.A. also acknowledges partial support from the Bulgarian NSF grants DFNI T02/6 and DN 08/3, as well as the bilateral grant STC/Bulgaria-France 01/6.

\appendix

\section{Reality conditions for unifying form of $H$} \label{HComplexConsts}
\setcounter{equation}{0}

Here we will show that the types of choices for the complex constants $h$ and $k$, that give a real function in (\ref{HNewSol}), are quite restricted and only lead to the solutions in (\ref{Hsol}). For that purpose, it is technically more convenient to consider the derivative $\dot{H} (t)$. Clearly, if the latter is a real function of $t$, then so is $H(t)$ itself. 

From (\ref{HNewSol}) we have that:
\be \label{dH}
\dot{H} (t) = - \frac{4 \,c \,k \,h^2}{( ke^{hct}-e^{-hct})^2} \,\,\, .
\ee
Let us study this expression for generic complex integration constants $h = h_1 + i h_2$ and $k = k_1 + i k_2$, where $h_1,h_2,k_1,k_2 \in \mathbb{R}$. Note that we keep the physical parameter $c$ real. Now, we want to find under what conditions the imaginary part of (\ref{dH}) cancels out. The expression for that imaginary part is rather cumbersome. So we will not write it down explicitly. However, its numerator has the following structure:
\be \label{ImagPart}
- 4 e^{-2h_1ct} c \left[ Co_1 \cos^2 (h_2ct) + Co_2 \cos (h_2ct) \sin (h_2ct) + Co_3 \right] \,\,\, , 
\ee 
where the coefficients $Co_i$ are:
\bea \label{Co123}
Co_1 &=& 2 \left[ 1 - (k_2^2 + k_1^2) e^{4 h_1 ct} \right] k_2 h_2^2 - 4 \left[ 1 + (k_1^2 + k_2^2) e^{4h_1 ct} \right] k_1 h_1 h_2 \nn \\
&-& 2 \left[ 1 - (k_1^2 + k_2^2) e^{4h_1 ct} \right] k_2 h_1^2 \,\, , \nn \\
Co_2 &=& 2 \left[ 1 - (k_2^2 + k_1^2) e^{4 h_1 ct} \right] k_1 h_2^2 + 4 \left[ 1 + (k_1^2 + k_2^2) e^{4h_1 ct} \right] k_2 h_1 h_2  \nn \\
&-& 2 \left[ 1 - (k_1^2 + k_2^2) e^{4h_1 ct} \right] k_1 h_1^2 \,\, , \nn \\
Co_3 &=& - \left[ 1 - (k_2^2 + k_1^2) e^{4 h_1 ct} \right] k_2 h_2^2 + 2 \left[ 1 + (k_1^2 + k_2^2) e^{4h_1 ct} \right] k_1 h_1 h_2 \nn \\
&+& \left[ 1 - (k_1^2 + k_2^2) e^{4h_1 ct} \right] k_2 h_1^2 + 4 (k_1^2 + k_2^2) e^{2h_1 ct} h_1 h_2 \,\, .
\eea
Since we want (\ref{ImagPart}) to vanish for any $t$, we need each of the three terms with coefficients $Co_i$, $i=1,2,3$ to vanish separately. Notice that $Co_1 + 2 Co_3$ gives:
\be \label{k1k2h1h2}
8 (k_1^2 + k_2^2) e^{2h_1 ct} h_1 h_2 \,\,\, .
\ee
The condition that (\ref{k1k2h1h2}) equals zero, for $\forall \,t$ and for $k \!= \!k_1 + i k_2 \neq 0$\,, can be solved only by taking either $h_1 = 0$ or $h_2 = 0$. 

In the case $h_2 = 0$, the term with the coefficient $Co_2$ in (\ref{ImagPart}) vanishes, because of the $\sin (0)$ there, despite $Co_2 \neq 0$. Substituting $h_2=0$ in (\ref{Co123}) and requiring $Co_1 = 0$ and $Co_3 = 0$, we then find the same condition from both: 
\be \label{constrk2}
\left[ 1 - (k_1^2 + k_2^2) e^{4h_1 ct} \right] k_2 h_1^2 = 0 \,\,\, .
\ee
Since we want $h \neq 0$, we must have $h_1 \neq 0$ here. Therefore, the constraint (\ref{constrk2}) can only be solved {\it for any $t$} by taking $k_2 = 0$. So we have shown that if $h$ is real, then $k$ is real as well.

Now, let us consider the other case, namely with $h_1 = 0$ and thus with a purely imaginary $h$. In this case, all $Co_i$, $i=1,2,3$ coefficients are proportional to the factor $\left[ 1-(k_1^2+k_2^2) \right]$, as can be seen from (\ref{Co123}). Then, since we want both $k\neq 0$ and $h\neq 0$, it is easy to realize that the only way to guarantee the vanishing of all three $Co_i$'s is to have:
\be
k_1^2 + k_2^2 = 1 \,\,\, .
\ee
Obviously, this can be solved by $k_1 = \pm \cos (\theta)$ and $k_2 = \pm \sin (\theta)$ with arbitrary $\theta$. In other words, the constant $k$ has to be a pure phase. So we have shown that, when $h$ is purely imaginary, we have $k = \pm e^{i \theta}$.

To summarize, we have proved that to have a real function in (\ref{dH}), and thus a real function $H(t)$ in (\ref{HNewSol}), the constant $h$ has to be either purely real or purely imaginary. In the first case, the constant $k$ has to be real as well, whereas in the second case it has to be of the form $k = \pm e^{i \theta}$.

\section{Properties of the complex {\it arctanh}} \label{ComplexArctanh}
\setcounter{equation}{0}

To investigate the inflaton solution, in the main text, we need to deal with the function ${\it arctanh}$ for arguments that are outside of the domain, in which it is a real function. Here we show that, despite that, its imaginary part is a constant in the cases of interest for us and thus can be removed due to the additive integration constant in the inflaton solution. We also derive useful expressions for the real part of the complex {\it arctanh} in these cases. 

Let us start by reviewing some necessary definitions. For a complex argument $z$, the functions {\it arctanh} and ${\it arccoth}$ are defined respectively as:
\be \label{arctC}
{\rm arctanh} (z) = \frac{1}{2} \left[ \ln (1+z) \!- \ln (1-z) \right]
\ee
and
\be \label{arccC}
{\rm arccoth} (z) = \frac{1}{2} \ln \!\left( \frac{z+1}{z-1} \right)
\ee
in terms of the complex logarithm. The definition of the latter is:
\be \label{lnC}
\ln \!z = \ln \!|z| + i Arg(z) \,\,\, .
\ee
Also, as usual, the complex conjugate of the last expression is:
\be \label{lnCC}
(\ln \!z)^* = \ln \!|z| - i Arg(z) \,\,\, .
\ee
Note that, for a complex argument, all of the above functions are multi-valued, as is well-known. Here we always mean their principal values.

From (\ref{arctC}) it is clear that ${\it arctanh} (z)$ is real when $z$ is real and $z\in(-1,1)$, while (\ref{arccC}) shows that ${\it arccoth} (z)$ is real for $z$ either in $(-\infty,-1)$ or in $(1,\infty)$. In the main text, however, we will need ${\it arctanh} (z)$ either for a real argument that is $z>1$, or for a complex argument of the form $z=e^{i \varphi}$ with $\varphi$ being some angular variable. So here we will investigate both of these cases in turn. 

\subsection{arctanh\!\! $(z)$ for $z>1$} \label{Apzge1}

In this subsection we will show that ${\it arctanh} (z)$ for any $z>1$ has a constant imaginary part and, in fact, is given by:
\be
{\rm arctanh} (z) = {\rm arccoth} (z) - i \frac{\pi}{2} \,\,\, .
\ee

Let us start by considering the imaginary part: 
\be
{\rm Im} \left( {\rm arctanh } (z) \right) = \frac{1}{2 i} \left[\,{\rm arctanh } (z) - ({\rm arctanh } (z))^*\,\right] \,\,\, .
\ee
Using (\ref{arctC}), (\ref{lnC}) and (\ref{lnCC}) and keeping in mind that, for $z>1$\,, only the logarithm \,${\it ln} \,(1-z)$ \,is complex, we find:
\be
{\rm Im} \left( {\rm arctanh } (z) \right) = - \frac{1}{4i} \left[ \,\ln (1-z) - \left( \ln(1-z) \right)^* \,\right] = - \,\frac{1}{2} \,Arg (1-z) \,\,\, .
\ee
Since, for any real $z>1$, the expression $(1-z)$ is always a real negative number, we have that $Arg (1-z) = \pi$, regardless of the value of $z$. Hence, we obtain:
\be \label{Imarctanhzg1}
{\rm Im} \!\left( {\rm arctanh} ( z ) \right) = - \frac{\pi}{2} \qquad {\rm for} \qquad z>1 \quad .
\ee
In the case of interest for us in Section \ref{SecInflaton}, we have $z = \sqrt{k} e^{hct}$ with $k>1$ and $h,c,t>0$. So, in particular, (\ref{Imarctanhzg1}) gives:
\be \label{Imarctanh_zge1}
{\rm Im} \!\left( {\rm arctanh} ( \sqrt{k} e^{hct}) \right) = - \frac{\pi}{2} \,\,\, .
\ee

Now, let us show that the real part of ${\it arctanh} (z)$ for $z>1$ is given by ${\it arccoth} (z)$. For that purpose, consider the expression:
\be
{\rm Re} \left( {\rm arctanh } (z) \right) = \frac{1}{2} \left[ {\rm arctanh } (z) + ({\rm arctanh } (z))^* \right] \,\, .
\ee
Again using (\ref{arctC}), (\ref{lnC}) and (\ref{lnCC}), we obtain:
\be \label{ReInterm}
{\rm Re} \left( {\rm arctanh } (z) \right) = \frac{1}{2} \left[ \,\ln (1+z) - \,\ln \!|1-z| \,\right] \,\, .
\ee
Since, for $z>1$ we always have $|1-z| = z-1$ regardless of the value of $z$, (\ref{ReInterm}) gives:
\be \label{Rearctharccth}
{\rm Re} \left( {\rm arctanh } (z) \right) = \frac{1}{2} \left[ \ln (1+z) - \ln (z - 1) \right] \,\, .
\ee
Now, due to (\ref{arccC}), the result (\ref{Rearctharccth}) immediately implies that:
\be \label{Rearctanh_zge1}
{\rm Re} \left( {\rm arctanh } (z) \right) = {\rm arccoth } (z) \qquad {\rm for} \qquad z>1 \quad .
\ee
Specifying for the particular case of interest in Section \ref{SecInflaton}, i.e. with $z = \sqrt{k} e^{hct} > 1$, we have:
\be \label{Rearctanh}
{\rm Re} \!\left( {\rm arctanh} ( \sqrt{k} e^{hct}) \right) = {\rm arccoth} ( \sqrt{k} e^{hct}) \,\,\, .
\ee

\subsection{arctanh\!\! $( e^{i \varphi})$ for $\varphi \in [0,\frac{\pi}{2}]$} \label{ApExpIphi_0}

In this subsection we will show that ${\it arctanh} ( e^{i \varphi} )$ also has a constant imaginary part and, in the process, we will find a useful expression for its real part.

In view of the definitions (\ref{arctC}) and (\ref{lnC}), let us begin by computing $|1 \pm z|$ and $Arg (1 \pm z)$ for $z = e^{i \varphi}$. We find:
\bea \label{AbsArg}
&&|1+z| = \sqrt{2 (1 + \cos \varphi)} \quad \,\, , \quad \,Arg (1+z) = \arctan \left( \frac{\sin \varphi}{1 + \cos \varphi} \right) \,\, , \nn \\
&&|1-z| = \sqrt{2 (1 - \cos \varphi)} \quad \,\, , \quad \,Arg (1-z) = - \arctan \left( \frac{\sin \varphi}{1 - \cos \varphi} \right) \,\, .
\eea
Note that one can rewrite the expressions inside the ${\it arctan}$ as:
\be \label{Rewrite_arctan}
\frac{\sin \varphi}{1 + \cos \varphi} = \tan \!\left( \frac{\varphi}{2} \right) \qquad {\rm and} \qquad \frac{\sin \varphi}{1 - \cos \varphi} = \cot \!\left( \frac{\varphi}{2} \right) \qquad .
\ee
Substituting (\ref{AbsArg})-(\ref{Rewrite_arctan}) in (\ref{arctC}) and (\ref{lnC}), we obtain:
\be \label{arctanh_phase_interm}
{\rm arctanh} \!\left( e^{i \varphi} \right) = \frac{1}{4} \ln \!\left( \frac{1+\cos \varphi}{1-\cos \varphi} \right) + \frac{i}{2} \left( \frac{\varphi}{2} + \arctan \!\left[ \cot \!\left( \frac{\varphi}{2} \right) \right] \right) \,\,\, .
\ee
The real part can be rewritten as:
\be \label{Repartsimp}
\frac{1}{4} \ln \!\left( \frac{1+\cos \varphi}{1-\cos \varphi} \right) = \frac{1}{2} \ln \!\left( \frac{\sin \varphi}{1 - \cos \varphi} \right) = \frac{1}{2} \ln \!\left[ \cot \!\left( \frac{\varphi}{2} \right) \right] \,\,\, .
\ee

To simplify the imaginary part of (\ref{arctanh_phase_interm}), recall that for any complex $w$ with ${\rm Re} \,w \ge 0$ one has \cite{AS}:
\be \label{arctarccotpi}
\arctan (w) + {\rm arccot} (w) = \frac{\pi}{2} \,\,\, ,
\ee
where we have assumed, as always, that we are working with the principal values of the inverse trigonometric functions. Otherwise, one would have to add $n \pi$, $n = 0, \pm 1, \pm 2,...$ to the right-hand side of (\ref{arctarccotpi}). In our case $w = \cot(\frac{\varphi}{2})$ and indeed $w>0$ for $\varphi \in [0,\frac{\pi}{2}]$\,. Note that, if we were considering $\varphi \in [\pi, \frac{3 \pi}{2}]$ instead, which is the other interval ensuring $H_{(3)}>0$ as discussed in Section \ref{SecInflaton}, then we would have $-\frac{\pi}{2}$ on the right-hand side of (\ref{arctarccotpi}).\footnote{This would only affect the precise value of the imaginary part in (\ref{arctanhphase_final}). Clearly, though, the latter is irrelevant; the only important point is that it is the same constant within each interval of interest.} Now, we can simplify: $\arctan \!\left[ \cot \!\left( \frac{\varphi}{2} \right) \right] = \frac{\pi}{2} - {\rm arccot} \!\left[ \cot \!\left( \frac{\varphi}{2} \right) \right] = \frac{\pi}{2} - \frac{\varphi}{2}$\,. Using this, together with (\ref{Repartsimp}), inside (\ref{arctanh_phase_interm}), we obtain:
\be \label{arctanhphase_final}
{\rm arctanh} \!\left( e^{i \varphi} \right) = \frac{1}{2} \ln \!\left[ \cot \!\left( \frac{\varphi}{2} \right) \right] + i \frac{\pi}{4} \,\,\, .
\ee

Finally, let us note that, while the function \,$\ln \!\left[ \,\cot \!\left( \frac{\varphi}{2} \right) \right]$ \,is real for $\varphi \in [0,\frac{\pi}{2}]$\,, it is complex in the other interval with $H_{(3)}>0$\,, namely $\varphi \in [\pi,\frac{3\pi}{2}]$\,, since \,$\cot \!\left( \frac{\varphi}{2} \right) \!\!< \!0$ \,there. To make the relation between the two explicit, let us consider \,$\ln \!\left[ \,\cot \!\left( \frac{\hat{\varphi}}{2} \right) \right]$ for $\hat{\varphi} \in [\pi,\frac{3\pi}{2}]$ and take $\hat{\varphi} = \varphi + \pi$ with $\varphi \in [0,\frac{\pi}{2}]$\,. Then, we have:
\be
\ln \!\left[ \cot \!\left( \frac{\hat{\varphi}}{2} \right) \right] = \ln \!\left[ \cot \!\left( \frac{\varphi + \pi}{2} \right) \right] = \ln \!\left[ - \tan \!\left( \frac{\varphi}{2} \right) \right] = - \ln \!\left[ \cot \!\left( \frac{\varphi}{2} \right) \right] + i \pi \,\, ,
\ee
where in the last equality we have used that $\ln (-1) = i \pi$ on the principle branch.

\subsection{arctanh\!\! $(e^{i \varphi})$ for $\varphi \in [\pi,\frac{3\pi}{2}]$} \label{ApExpIphi}

In Section \ref{SecInflaton} we saw that the inflaton solution for case $(4)$ is given by the function \,${\rm arctanh} (i e^{i \varphi})$\,. $\!\!$Of course, for $\varphi \in [\frac{\pi}{2},\pi]$\,, this is just \,${\rm arctanh} (e^{i \hat{\varphi}})$ \,with $\hat{\varphi} \in [\pi,\frac{3\pi}{2}]$\,. Despite that, the form of the solution $\phi_{(4)}$ in (\ref{phicases}), which is in agreement with \cite{MSY}, looks very different from the right-hand side of (\ref{arctanhphase_final}). Here we will show that they are indeed the same.

Let us first note that the function \,${\rm arctanh} \!\left[\tan (\frac{\varphi}{2})\right]$ \,in (\ref{phicases}) is actually complex for $\varphi \in [\frac{\pi}{2},\pi]$ as \,$\tan (\frac{\varphi}{2}) \!> \!1$ \,in that interval; see Appendix \ref{Apzge1}. So it is more convenient to work in the other interval with $H_{(4)}>0$\,, namely $\varphi \in [\frac{3\pi}{2},2\pi]$\,, since \,${\rm arctanh} \!\left[\tan (\frac{\varphi}{2})\right]$ \,is real there. We will see that, as should be expected, the real part is always the same (up to a sign) and the imaginary part is always constant in every interval of interest.\footnote{Plus, of course, the actual value of the imaginary part is irrelevant due to presence of the additive integration constant $C^{\phi}_4$.}

Now, let us consider the general expression ${\it arctanh} (i e^{i \varphi})$, that we found in (\ref{case4IexpIVarphi}) for case $(4)$, inside the interval $\varphi \in [\frac{3\pi}{2},2\pi]$. It is straightforward, albeit quite tedious, to show that one obtains:
\be \label{arctanhiphase_final}
{\rm arctanh} \!\left( i e^{i \varphi} \right) = - \,{\rm arctanh} \!\left[ \tan \!\left( \frac{\varphi}{2} \right) \right] + i \frac{\pi}{4} \,\,\, ,
\ee
by repeating the same kind of steps as in Appendix \ref{ApExpIphi_0}, namely computing $|1\pm z|$ and $Arg (1\pm z)$ for $z=i e^{i \varphi}$ etc.. So we will not write down this calculation here. Let us just note that the only new ingredient one needs is that $\arctan (\frac{1}{w}) = {\rm arccot} (w)$ for any complex $w$ .\footnote{It is also useful to note the relations \,\,$\sin \varphi = \frac{2 \tan \left( \frac{\varphi}{2} \right)}{1 + \tan^2 \!\left( \frac{\varphi}{2} \right)}$ \,and \,$\cos \varphi = \frac{1 - \tan^2 \!\left(\frac{\varphi}{2} \right)}{1 + \tan^2 \!\left( \frac{\varphi}{2} \right)}$\,\,\,.}

The results (\ref{arctanhphase_final}) and (\ref{arctanhiphase_final}) already show that \,$\frac{1}{2} \ln \!\left[ \,\cot \!\left( \frac{\varphi}{2} \right) \right]$ \,and \,${\rm arctanh} \!\left[ \tan \!\left( \frac{\varphi}{2} \right) \right]$ \,give the same function in the corresponding intervals, since obviously ${\rm arctanh} (e^{i \varphi})$ for $\varphi \in [0,\frac{\pi}{2}]$ is the same as ${\rm arctanh} (i e^{i \varphi})$ for $\varphi \in [\frac{3\pi}{2},2\pi]$. Nevertheless, we can see more directly how the real functions in the two cases transform into each other in the following way. Let us consider \,${\rm arctanh} \!\left[ \tan \!\left( \frac{\hat{\varphi}}{2} \right) \right]$ for $\hat{\varphi} \in [\frac{3\pi}{2},2\pi]$ and write $\hat{\varphi} = \varphi + \frac{3\pi}{2}$ with $\varphi \in [0,\frac{\pi}{2}]$. Then, we obtain:
\bea
{\rm arctanh} \!\left[ \tan \!\left( \frac{\hat{\varphi}}{2} \right) \right] \!&=& \!- \,{\rm arctanh} \!\left[ \cot \!\left( \frac{\varphi + \frac{\pi}{2}}{2} \right) \right] = - \frac{1}{2} \ln \!\!\left[ \frac{1 + \cot \left( \frac{\varphi + \frac{\pi}{2}}{2} \right)}{1-\cot \left( \frac{\varphi + \frac{\pi}{2}}{2} \right)} \right] \nn \\
\!&=& \!- \frac{1}{2} \ln \!\left[ \frac{\cos \varphi + 1 -\sin \varphi}{\cos \varphi -1 + \sin \varphi} \right] = - \frac{1}{2} \ln \!\left[ \cot \!\left( \frac{\varphi}{2} \right) \right] \,\, ,
\eea
where in the intermediate steps we have used that $\cot \!\left( \frac{\alpha}{2} \right) = \frac{1+\cos \alpha}{\sin \alpha}$ and $\cot \!\left( \frac{\alpha}{2} \right) = \frac{\sin \alpha}{1 - \cos \alpha}$ for any $\alpha$.

To complete the proof that the inflaton solutions $\phi_{(3)} (t)$ and $\phi_{(4)} (t)$ in (\ref{phicases}) give the same function, let us also consider \,${\rm arctanh} \!\left[ \tan \!\left( \frac{\hat{\varphi}}{2} \right) \right]$ \,for $\hat{\varphi} \in [\frac{\pi}{2},\pi]$. Taking $\hat{\varphi} = \varphi + \frac{\pi}{2}$ with $\varphi \in [0,\frac{\pi}{2}]$, we have:
\bea \label{EqApB3}
{\rm arctanh} \!\left[ \tan \!\left( \frac{\hat{\varphi}}{2} \right) \right] \!\!\!&=& \!\!\!- \,{\rm arctanh} \!\left[ \cot \!\left( \frac{\varphi}{2} - \frac{\pi}{4} \right) \right] =  - \frac{1}{2} \ln \!\!\left[ \frac{1+ \cot \left( \frac{ \varphi - \frac{\pi}{2} }{2} \right) }{1- \cot \left (\frac {\varphi - \frac{\pi}{2} }{2} \right)} \right] \\
\!\!\!&=& \!\!\!- \frac{1}{2} \ln \!\!\left[ \frac{1 - \frac{1+\sin \varphi}{\cos \varphi}}{1 + \frac{1+\sin \varphi}{\cos \varphi}} \right] \!= - \frac{1}{2} \ln \!\!\left[ \frac{-1}{ \cot \!\left( \frac{\varphi}{2} \right)} \right]  \!=  \frac{1}{2} \ln \!\left[ \cot \!\left( \frac{\varphi}{2} \right) \right] \!- i \frac{\pi}{2} \,\,\, , \nn
\eea
where we have used (\ref{Rewrite_arctan}) multiple times. Also, at the last step we used that $\ln (-1) = i \pi$ on the principle branch. 

Given the overall $\pm$ in front of the inflaton function, as well as the additive integration constant, the results in this Appendix show that indeed $\phi_{(3)}$ and $\phi_{(4)}$ in (\ref{phicases}) give the same solution. Also, (\ref{EqApB3}) makes explicit the fact that ${\rm arctanh} \!\left[ \tan \!\left( \frac{\hat{\varphi}}{2} \right) \right]$ is complex for $\hat{\varphi} \in [\frac{\pi}{2},\pi]$ with imaginary part, that is in complete agreement with Appendix \ref{Apzge1}.


\begin{thebibliography}{100}

\bibitem{CMB}
C. Bennett, A. Banday, K. Gorski, G. Hinshaw, P. Jackson, et al., {\em Four year COBE DMR
cosmic microwave background observations: Maps and basic results}, Astrophys.J. {\bf 464} (1996) L1, astro-ph/9601067;
{\bf WMAP} Collaboration, C. Bennett et al., {\em First year Wilkinson Microwave Anisotropy Probe (WMAP) observations: Preliminary maps and basic results}, Astrophys.J.Suppl. {\bf 148} (2003) 1, astro-ph/0302207; {\bf WMAP} Collaboration, D. Spergel et al., {\em First year Wilkinson Microwave Anisotropy Probe (WMAP) observations: Determination of cosmological parameters}, Astrophys.J.Suppl. {\bf 148} (2003) 175, astro-ph/0302209;
{\bf Planck} Collaboration, P. Ade et al., {\em Planck 2013 results. XV. CMB power spectra and likelihood}, Astron.Astrophys. {\bf 571} (2014) A15, arXiv:1303.5075 [astro-ph.CO]; {\bf Planck} Collaboration, P. Ade et al., {\em Planck 2015 results. XX. Constraints on inflation}, Astron. Astrophys. {\bf 594} (2016) A20, arXiv:1502.02114 [astro-ph.CO].

\bibitem{TW}
N. Tsamis and R. Woodard, {\em Improved Estimates of Cosmological Perturbations}, Phys. Rev. {\bf D69} (2004) 084005, astro-ph/0307463.

\bibitem{WK}
W. Kinney, {\em Horizon crossing and inflation with large $\eta$}, Phys. Rev. {\bf D72} (2005)
023515, gr-qc/0503017.

\bibitem{ultra-slow-roll}
J. Martin, H. Motohashi and T. Suyama, {\em Ultra Slow-Roll Inflation and the non-Gaussianity Consistency Relation}, Phys. Rev. {\bf D87} (2013) 023514, arXiv:1211.0083 [astro-ph.CO]; M. H. Namjoo, H. Firouzjahi and M. Sasaki, {\em Violation of non-Gaussianity consistency relation in a single field inflationary model}, Europhys. Lett. {\bf 101} (2013) 39001, arXiv:1210.3692 [astro-ph.CO]; S. Mooij and G. A. Palma, {\em Consistently violating the non-Gaussian consistency relation}, JCAP {\bf 1511} (2015) 025, arXiv:1502.03458 [astro-ph.CO]; S. Hirano, T. Kobayashi and S. Yokoyama, {\em Ultra slow-roll G-inflation}, Phys. Rev. {\bf D94} (2016) 103515, arXiv:1604.00141 [astro-ph.CO]; A. E. Romano, S. Mooij and M. Sasaki, {\em Global adiabaticity and non-Gaussianity consistency condition}, Phys. Lett. {\bf B761} (2016) 119, arXiv:1606.04906 [gr-qc]; C. Germani and T. Prokopec, {\em On primordial black holes from an inflection point}, Phys. Dark Univ. {\bf 18} (2017) 6, arXiv:1706.04226 [astro-ph.CO]; K. Dimopoulos, {\em Ultra slow-roll inflation demystified}, arXiv:1707.05644 [hep-ph].

\bibitem{CK}
J. Cook and L. Krauss, {\em Large Slow Roll Parameters in Single Field Inflation}, JCAP
{\bf 1603} (2016) 028, arXiv:1508.03647 [astro-ph.CO].

\bibitem{MSY}
H. Motohashi, A. Starobinsky, J. Yokoyama, {\em Inflation with a constant rate of roll}, JCAP {\bf 1509} (2015) 018, arXiv:1411.5021 [astro-ph.CO].

\bibitem{MS}
H. Motohashi, A. Starobinsky, {\em Constant-roll inflation: confrontation with recent observational data}, Europhys. Lett. {\bf 117} (2017) 39001, arXiv:1702.05847 [astro-ph.CO].

\bibitem{OO}
S.D. Odintsov, V.K. Oikonomou, {\em Inflationary Dynamics with a Smooth Slow-Roll to Constant-Roll Era Transition}, JCAP {\bf 1704} (2017) 041, arXiv:1703.02853 [gr-qc]; {\em Inflation with a Smooth Constant-Roll to Constant-Roll Era Transition}, Phys.Rev. {\bf D96} (2017) 024029, arXiv:1704.02931 [gr-qc].

\bibitem{VO}
V. Oikonomou, {\em A Smooth Constant-Roll to a Slow-Roll Modular Inflation Transition}, arXiv:1709.02986 [gr-qc].

\bibitem{NOO}
S. Nojiri, S. Odintsov and Oikonomou, {\em Constant-roll Inflation in $F(R)$ Gravity}, arXiv:1704.05945 [gr-qc].

\bibitem{MS2}
H. Motohashi and A. Starobinsky, {\em $f(R)$ constant-roll inflation}, Eur.Phys.J. {\bf C77} 538 (2017), arXiv:1704.08188 [astro-ph.CO].

\bibitem{VO2}
V. Oikonomou, {\em Reheating in Constant-roll $F(R)$ Gravity}, Mod.Phys.Lett. {\bf A32} (2017) 1750172, arXiv:1706.00507 [gr-qc].

\bibitem{OOS}
S. Odintsov, V. Oikonomou and L. Sebastiani, {\em Unification of Constant-roll Inflation and Dark Energy with Logarithmic $R^2$-corrected and Exponential $F(R)$ Gravity}, Nucl.Phys. {\bf B923} (2017) 608, arXiv:1708.08346 [gr-qc].

\bibitem{AHNOO}
A. Awad, W. El Hanafy, G. Nashed, S. Odintsov and V. Oikonomou, {\em Constant-roll Inflation in $f(T)$ Teleparallel Gravity}, arXiv:1710.00682 [gr-qc].

\bibitem{LPB}
A. Liddle, P. Parsons, J. Barrow, {\em Formalising the Slow-Roll Approximation in Inflation}, Phys.Rev. {\bf D50} (1994) 7222, astro-ph/9408015.

\bibitem{DB}
D. Baumann, {\em TASI Lectures on Inflation}, Conference Proceedings for ``Physics of the Large and the Small", Theoretical Advanced Study Institute in Elementary Particle Physics, Boulder, Colorado, June 2009, arXiv:0907.5424 [hep-th].

\bibitem{AW}
L. Abbott and M. Wise, {\em Constraints on Generalized Inflationary Cosmologies}, Nucl. Phys. {\bf B244} (1984) 541.

\bibitem{LM}
F. Lucchin and S. Matarrese, {\em Power Law Inflation}, Phys. Rev. {\bf D32} (1985) 1316.

\bibitem{AS}
M. Abramowitz and I. Stegun, {\em Handbook of Mathematical Functions} (Dover Publications, Inc., New York, 1972).

\bibitem{VM}
V. Mukhanov, {\em Gravitational Instability of the Universe Filled with a Scalar Field}, JETP Lett. {\bf 41} (1985) 493.

\bibitem{MSa}
M. Sasaki, {\em Large Scale Quantum Fluctuations in the Inflationary Universe}, Prog. Theor. Phys. {\bf 76} (1986) 1036.

\bibitem{LL}
D. Lyth and A. Liddle, {\em The Primordial Density Perturbation: Cosmology, Inflation and the Origin of Structure} (Cambridge University Press, 2009).

\bibitem{EFTofI}
C. Cheung, P. Creminelli, A.L. Fitzpatrick, J. Kaplan, L. Senatore, {\em The Effective Field Theory of Inflation}, JHEP {\bf 0803} (2008) 014, arXiv:0709.0293 [hep-th].

\bibitem{MH}
H. Motohashi and W. Hu, {\em Generalized Slow Roll in the Unified Effective Field Theory of Inflation}, Phys. Rev. {\bf D96} (2017) 023502, arXiv:1704.01128 [hep-th].

\bibitem{CJS}
P. Channuie, J. Joergensen and F. Sannino, {\em Minimal Composite Inflation}, JCAP
{\bf 1105} (2001) 007, arXiv:1102.2898 [hep-ph].

\bibitem{BCJS}
F. Bezrukov, P. Channuie, J. Joergensen and F. Sannino, {\em Composite Inflation Setup
and Glueball Inflation}, Phys. Rev. {\bf D86} (2012) 063513, arXiv:1112.4054 [hep-ph].

\bibitem{ChK}
P. Channuie and K. Karwan, {\em Large Tensor-to-Scalar Ratio from Composite Inflation}, Phys. Rev. {\bf D90} (2014) 047303, arXiv:1404.5879 [astro-ph.CO].

\bibitem{LA}
L. Anguelova, {\em A Gravity Dual of Ultra-slow Roll Inflation}, Nucl. Phys. {\bf B911} (2016) 480, arXiv:1512.08556 [hep-th]. 

\bibitem{ASW}
L. Anguelova, P. Suranyi and L.C.R. Wijewardhana, {\em De Sitter Space in Gauge/Gravity Duality}, Nucl. Phys. {\bf B899} (2015) 651, arXiv:1412.8422 [hep-th]; {\em Toward a Gravity Dual of Glueball Inflation}, Bulg. J. Phys. {\bf 42} (2015) 277, arXiv:1507.04053 [hep-th].

\bibitem{LA2}
L. Anguelova, {\em On Slow-roll Glueball Inflation from Holography}, Bulg. J. Phys. {\bf 44} (2017) 48, arXiv:1611.00295 [hep-th].

\bibitem{LA3}
L. Anguelova, {\em Glueball Inflation and Gauge/Gravity Duality}, Springer Proc. Math. Stat. {\bf 191} (2016) 285, arXiv:1601.02449 [hep-th].

\end{thebibliography}
\end{document}